\documentclass[11pt, a4paper, notitlepage]{article}

\usepackage{amsmath, latexsym, amsthm, amsfonts}
\usepackage{graphicx}
\usepackage{epsfig}
\usepackage{varioref}
\usepackage{natbib} 

\bibpunct{(}{)}{;}{a}{}{,}

\theoremstyle{plain}
\newtheorem{thm}{Theorem}[section]
\newtheorem{cnd}{Condition}[section]

\newcommand{\infint}{\int_{-\infty}^{\infty}}
\def\convd{\stackrel{\cal D}{\rightarrow}}
\def\ex{{\rm E\,}}
\def\var{\mathop{\rm Var}\nolimits}

\begin{document}
\title{Some thoughts on the asymptotics of the deconvolution kernel density estimator}
\author{Bert van Es\\
{\normalsize Korteweg-de Vries Instituut voor Wiskunde}\\
{\normalsize Universiteit van Amsterdam}\\
{\normalsize Plantage Muidergracht 24}\\
{\normalsize 1018 TV Amsterdam}\\
{\normalsize The Netherlands}\\
{\normalsize vanes@science.uva.nl}\\
{}\\
{\normalsize Shota Gugushvili\footnote{The research of this author
was financially supported by the Nederlandse Organisatie voor
Wetenschappelijk Onderzoek (NWO). Part of the work was done while
this author was at the Korteweg-de Vries Instituut voor Wiskunde
in Amsterdam.}}\\
{\normalsize Eurandom}\\
{\normalsize Technische Universiteit Eindhoven}\\
{\normalsize P.O.\ Box 513}\\
{\normalsize 5600 MB Eindhoven}\\
{\normalsize The Netherlands}\\
{\normalsize gugushvili@eurandom.tue.nl}
}
\maketitle
\begin{abstract}
Via a simulation study we compare the finite sample performance of
the deconvolution kernel density estimator in the supersmooth
deconvolution problem to its asymptotic behaviour predicted by two
asymptotic normality theorems. Our results indicate that for lower
noise levels and moderate sample sizes the match between the
asymptotic theory and the finite sample performance of the
estimator is not satisfactory. On the other hand we show that the
two approaches produce reasonably close results for higher noise
levels. These observations in turn provide additional motivation
for the study of deconvolution problems under the assumption that
the error term variance $\sigma^2\rightarrow 0$ as the sample size
$n\rightarrow\infty.$
\end{abstract}
\newpage

\section{Introduction}

Let $X_1,\ldots,X_n$ be i.i.d.\ observations, where $X_i=Y_i+Z_i$
and the $Y$'s and $Z$'s are independent. Assume that the $Y$'s are
unobservable and that they have the density $f$ and also that the
$Z$'s have a known density $k.$ The deconvolution problem consists
in estimation of the density $f$ based on the sample
$X_1,\ldots,X_n.$

A popular estimator of $f$ is the deconvolution kernel density
estimator, which is constructed via Fourier inversion and kernel
smoothing. Let $w$ be a kernel function and $h>0$ a bandwidth. The
kernel deconvolution density estimator $f_{nh}$ is defined as
\begin{equation}
\label{fnh} f_{nh}(x)=\frac{1}{2\pi}\infint
e^{-itx}\frac{\phi_w(ht)\phi_{emp}(t)}{\phi_k(t)}dt=\frac{1}{nh}\sum_{j=1}^n w_h\left(\frac{x-X_j}{h}\right),
\end{equation}
where $\phi_{emp}$ denotes the empirical characteristic function of the sample, i.e.
\begin{equation*}
\phi_{emp}(t)=\frac{1}{n}\sum_{j=1}^n e^{itX_j},
\end{equation*}
$\phi_w$ and $\phi_k$ are Fourier transforms of the functions $w$
and $k,$ respectively, and
\begin{equation*}
w_h(x)=\frac{1}{2\pi}\infint e^{-itx}\frac{\phi_w(t)}{\phi_k(t/h)}dt.
\end{equation*}
The estimator \eqref{fnh} was proposed in \citet{carroll1} and \citet{carroll2} and there is a vast amount of
literature dedicated to it (for additional bibliographic information see e.g.\ \citet{vanes2} and \citet{vanes1}).

Depending on the rate of decay of the characteristic function $\phi_k$ at plus and minus infinity, deconvolution problems are usually divided into two groups, ordinary smooth deconvolution problems and supersmooth deconvolution problems. In the first case it is assumed that $\phi_k$ decays algebraically and in the second case the decay is essentially exponential. This rate of decay, and consequently the smoothness of the density $k,$ has a decisive influence on the performance of \eqref{fnh}. The general picture that one sees is that smoother $k$ is, the harder the estimation of $f$ becomes, see e.g.\ \citet{fan1}.

Asymptotic normality of \eqref{fnh} in the ordinary smooth case
was established in \citet{fan2}, see also \citet{fan4}. The limit
behaviour in this case is essentially the same as that of a kernel
estimator of a higher order derivative of a density. This is
obvious in certain relatively simple cases where the estimator is
actually equal to the sum of derivatives of a kernel density
estimator, cf.\ \citet{vanes0}.

Our main interest, however, lies in asymptotic normality of \eqref{fnh} in the supersmooth case. In this case under certain
conditions on the kernel $w$ and the unknown density $f,$ the following theorem was proved in
\citet{fan3}.
\begin{thm}
\label{thmanfan}
Let $f_{nh}$ be defined by \eqref{fnh}. Then
\begin{equation}
\label{anfan}
\frac{\sqrt{n}}{s_n}(f_{nh}(x)-\ex[f_{nh}(x)])\convd {\mathcal N}(0,1)
\end{equation}
as $n\rightarrow\infty.$ Here either $s_n^2=(1/n)\sum_{j=1}^n Z_{nj}^2,$ or $s_n^2$ is the sample variance of $Z_{n1},\ldots,Z_{nn}$ with
$Z_{nj}=(1/h)w_h((x-X_j)/h).$
\end{thm}

The asymptotic variance of $f_{nh}$ itself does not follow from this result.
On the other hand \citet{vanes2}, see also \citet{vanes1}, derived a central limit theorem for \eqref{fnh} where the
normalisation is deterministic and the asymptotic variance is given.

For the purposes of the present work it is sufficient to use the result of \citet{vanes1}. However, before recalling the corresponding theorem, we first formulate conditions on the kernel $w$ and the density $k.$
\begin{cnd}
\label{condw}
Let $\phi_w$ be real-valued, symmetric and have support $[-1,1].$ Let $\phi_w(0)=1,$ and assume $\phi_w(1-t)=At^{\alpha}+o(t^{\alpha})$ as $t\downarrow 0$ for some constants $A$ and $\alpha\geq 0.$
\end{cnd}
The simplest example of such a kernel is the sinc kernel
\begin{equation}
\label{sinckernel}
w(x)=\frac{\sin x}{\pi x}.
\end{equation}
Its characteristic function equals $\phi_{w}(t)=1_{[-1,1]}(t).$ In this case $A=1$ and $\alpha=0.$

Another kernel satisfying Condition \ref{condw} is
\begin{equation}
\label{fankernel}
w(x)=\frac{48\cos x}{\pi x^4}\left(1-\frac{15}{x^2}\right)-\frac{144\sin x}{\pi x^5}\left(2-\frac{5}{x^2}\right).
\end{equation}
Its corresponding Fourier transform is given by $\phi_{w}(t)=(1-t^2)^3 1_{[-1,1]}(t).$ Here $A=8$ and $\alpha=3.$ The kernel \eqref{fankernel}
was used for simulations in \citet{fan3} and its good performance in deconvolution context was established in
\citet{delaigle1}.

Yet another example is
\begin{equation}
\label{wandkernel}
w(x)=\frac{3}{8\pi}\left(\frac{\sin (x/4)}{x/4}\right)^4.
\end{equation}
The corresponding Fourier transform equals
\begin{equation*}
\phi_{w}(t)=2(1-|t|)^3 1_{[1/2,1]}(|t|)+(6|t|^3-6t^2+1)1_{[-1/2,1/2]}(t).
\end{equation*}
Here $A=2$ and $\alpha=3.$ This kernel was considered in \citet{wand} and \citet{delaigle1}.

Now we formulate the condition on the density $k.$

\begin{cnd}
\label{condk} Assume that $\phi_k(t)\sim C
|t|^{\lambda_0}\exp\left[-|t|^{\lambda}/\mu\right]$ as
$|t|\rightarrow \infty,$ for some $\lambda>1,\mu>0,\lambda_0$ and
some constant $C.$ Furthermore, let $\phi_k(t)\neq 0$ for all
$t\in{\mathbb{R}}.$
\end{cnd}

The following theorem holds true, see \citet{vanes1}.

\begin{thm}
\label{thman}
Assume Conditions \ref{condw} and \ref{condk} and let $\ex [X^2]<\infty.$ Then, as $n\rightarrow\infty$ and $h\rightarrow 0,$
\begin{equation*}
\frac{\sqrt{n}}{h^{\lambda(1+\alpha)+\lambda_0-1} e^{{1}/{(\mu
h^\lambda})}}\,(f_{nh}(x)-\ex [f_{nh}(x)])\convd {\mathcal
N}\left(0,\frac{A^2}{2\pi^2}\left(\frac{\mu}{\lambda}\right)^{2+2\alpha}(\Gamma(\alpha+1))^2\right).
\end{equation*}
Here $\Gamma$ denotes the gamma function.
\end{thm}

The goal of the present note is to compare the theoretical
behaviour of the estimator \eqref{fnh} predicted by Theorem
\ref{thman} to its behaviour in practice, which will be done via a
limited simulation study. The obtained results can be used to
compare Theorem \ref{thmanfan} to Theorem \ref{thman}, e.g.\
whether it is preferable to use the sample standard deviation
$s_n$ in the construction of pointwise confidence intervals
(computation of $s_n$ is more involved) or to use the
normalisation of Theorem \ref{thman} (this involves evaluation of
a simpler expression). The rest of the paper is organised as
follows: in Section \ref{simulations} we present some simulation results, while in
Section \ref{conclusions} we discuss the obtained results and draw conclusions.

\section{Simulation results}
\label{simulations}

All the simulations in this section were done in Mathematica. We considered three target densities. These densities are:
\begin{enumerate}
\item density \# 1: $Y\sim {\mathcal N}(0,1);$
\item density \# 2: $Y\sim \chi^2(3);$
\item density \# 3: $Y\sim 0.6{\mathcal N}(-2,1)+0.4{\mathcal N}(2,0.8^2).$
\end{enumerate}
The density \# 2 was chosen because it is skewed, while the
density \# 3 was selected because it has two unequal modes. We also
assumed that the noise term $Z$ was ${\mathcal N}(0,0.4^2)$
distributed. Notice that the noise-to-signal ratio
$\operatorname{NSR}=\var[Z]/\var[Y] 100\%$ for the density \# 1
equals $16\%,$ for the density \# 2 it is equal to $2.66\%,$ and
for the density \# 3 it is given by $3\%.$ We have chosen the
sample size $n=50$ and generated $500$ samples from the density
$g=f\ast k.$ Notice that such $n$ was also used in simulations in
e.g.\ \citet{delaigle3}. Even though at the first sight $n=50$
might look too small for  normal deconvolution,  for the low noise
level that we have the deconvolution kernel density estimator will
still perform well, cf.\ \citet{wand}. As a kernel we took the
kernel \eqref{fankernel}. For each model that we considered, the
theoretically optimal bandwidth, i.e.\ the bandwidth minimising
\begin{equation}
\label{mise}
\operatorname{MISE}[f_{nh}]=\ex \left[\infint (f_{nh}(x)-f(x))^2dx \right],
\end{equation}
the mean-squared error of the estimator $f_{nh},$ was selected by
evaluating \eqref{mise} for a grid of values of
$h_k=0.01k,k=1,\ldots,100,$ and selecting the $h$ that minimised
$\operatorname{MISE}[f_{nh}]$ on that grid. Notice that it is
easier to evaluate \eqref{mise} by rewriting it in terms of the
characteristic functions, which can be done via Parseval's
identity, cf.\ \citet{carroll2}. For  real data of course the
above method does not work, because \eqref{mise} depends on the
unknown $f.$ We refer to \citet{delaigle3} for  data-dependent
bandwidth selection methods in kernel deconvolution.

\begin{figure}[htb]
\setlength{\unitlength}{1cm}
\begin{minipage}{5.5cm}
\begin{picture}(5.5,4.0)
\epsfxsize=5.5cm\epsfysize=4cm\epsfbox{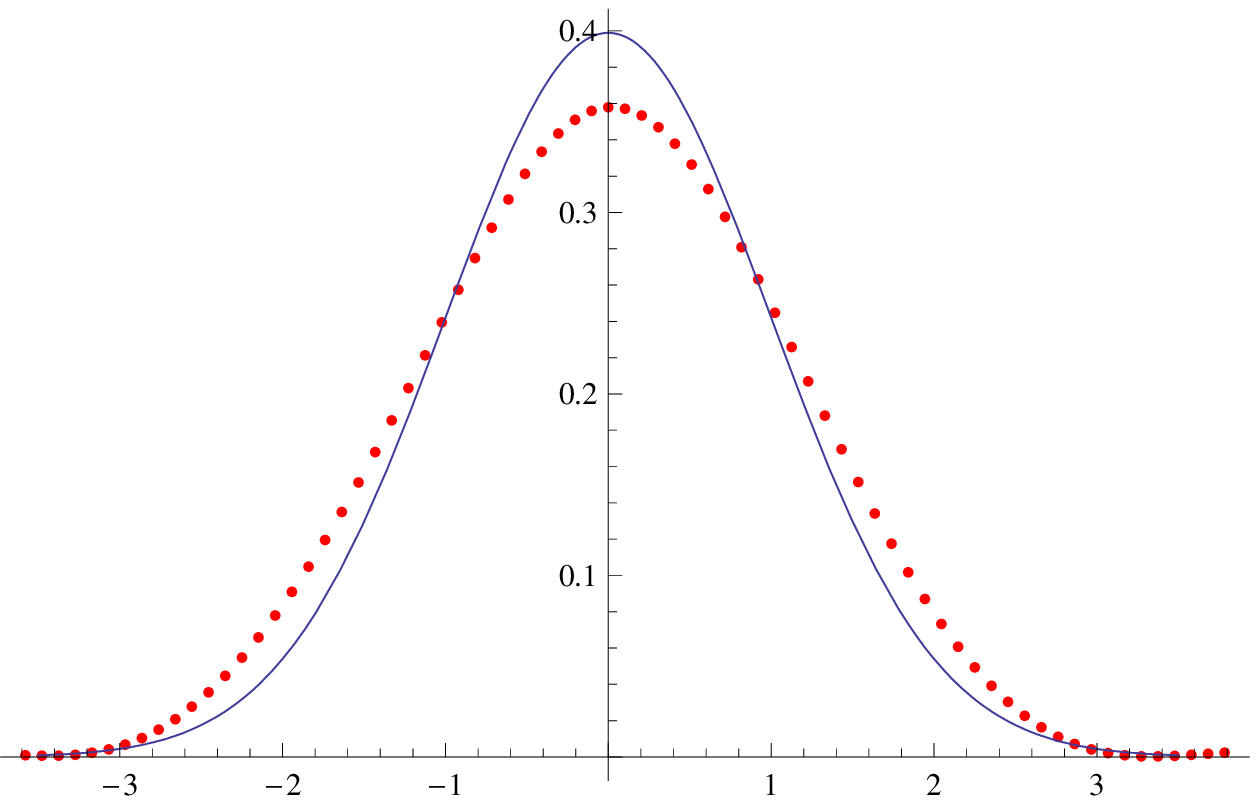}
\end{picture}
\end{minipage}
\hfill
\begin{minipage}{5.5cm}
\begin{picture}(5.5,4.0)
\epsfxsize=5.5cm\epsfysize=4cm\epsfbox{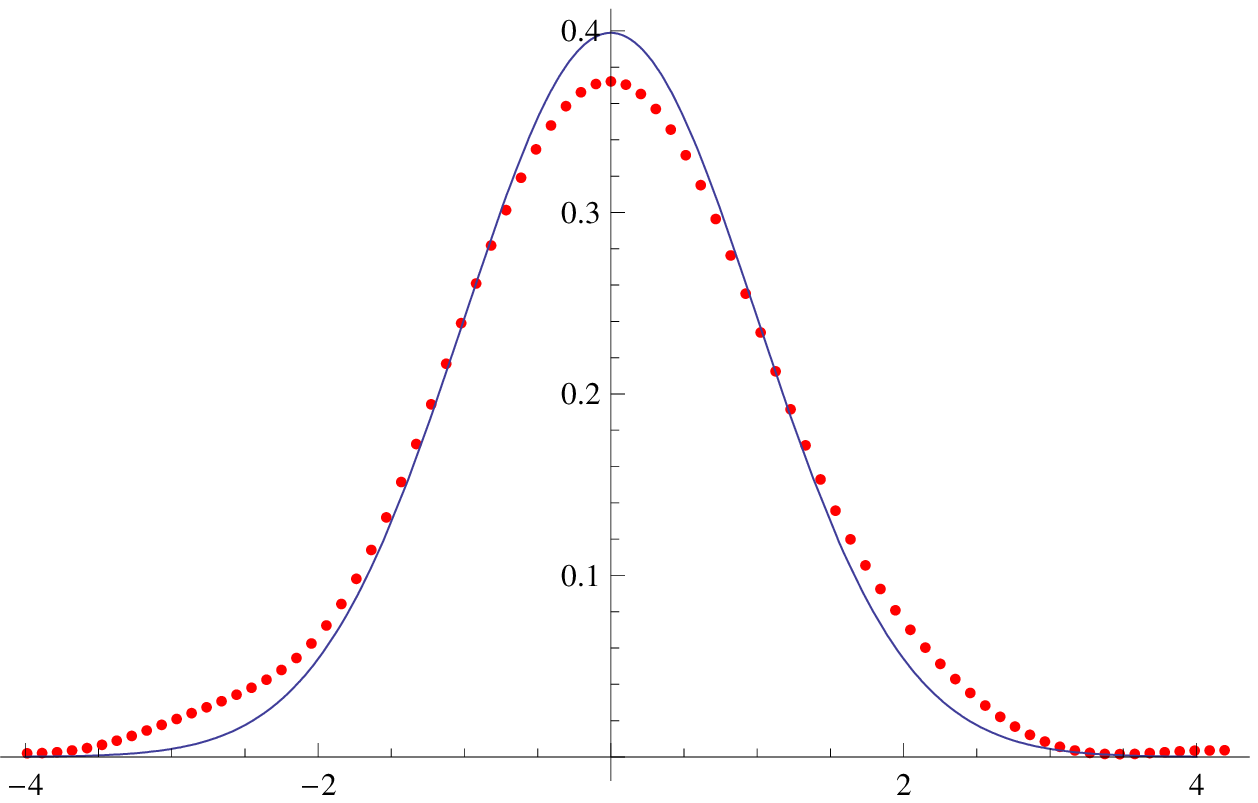}
\end{picture}
\end{minipage}
\vfill
\setlength{\unitlength}{1cm}
\begin{minipage}{5.5cm}
\begin{picture}(5.5,4.0)
\epsfxsize=5.5cm\epsfysize=4cm\epsfbox{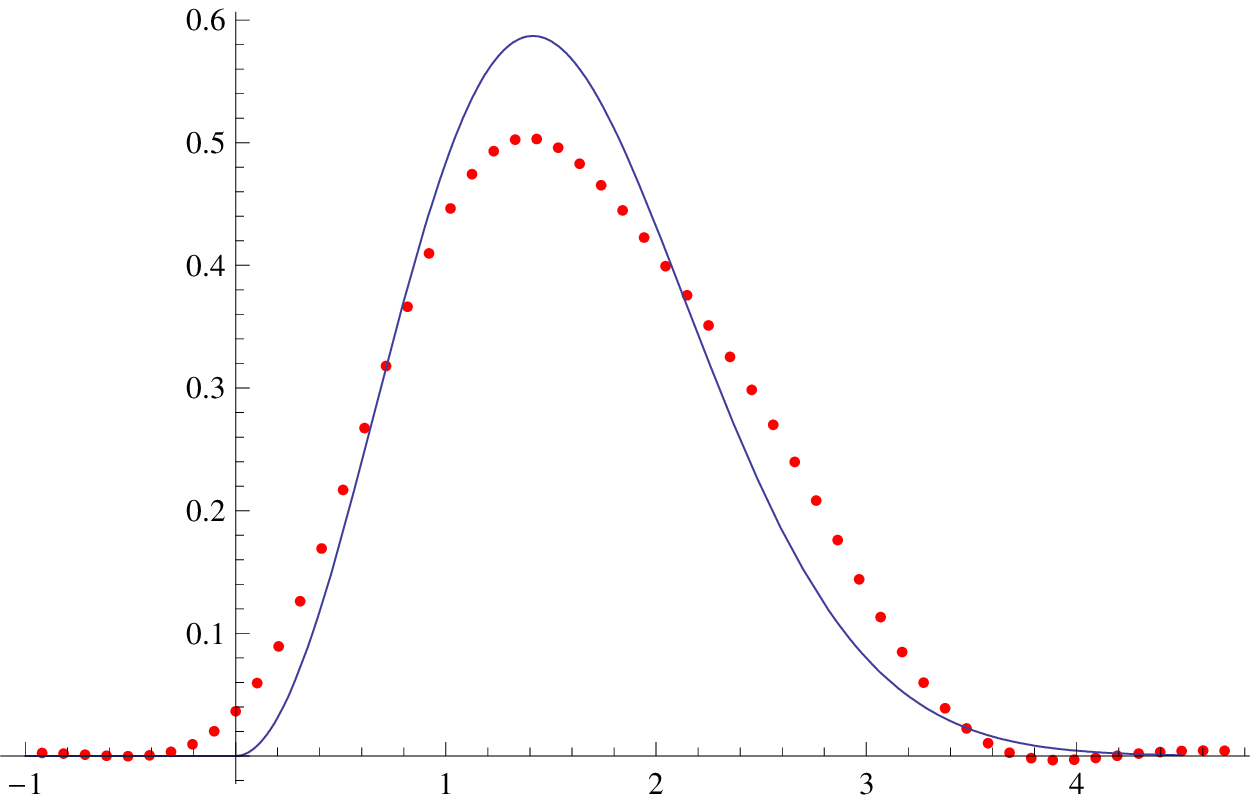}
\end{picture}
\end{minipage}
\hfill
\begin{minipage}{5.5cm}
\begin{picture}(5.5,4.0)
\epsfxsize=5.5cm\epsfysize=4cm\epsfbox{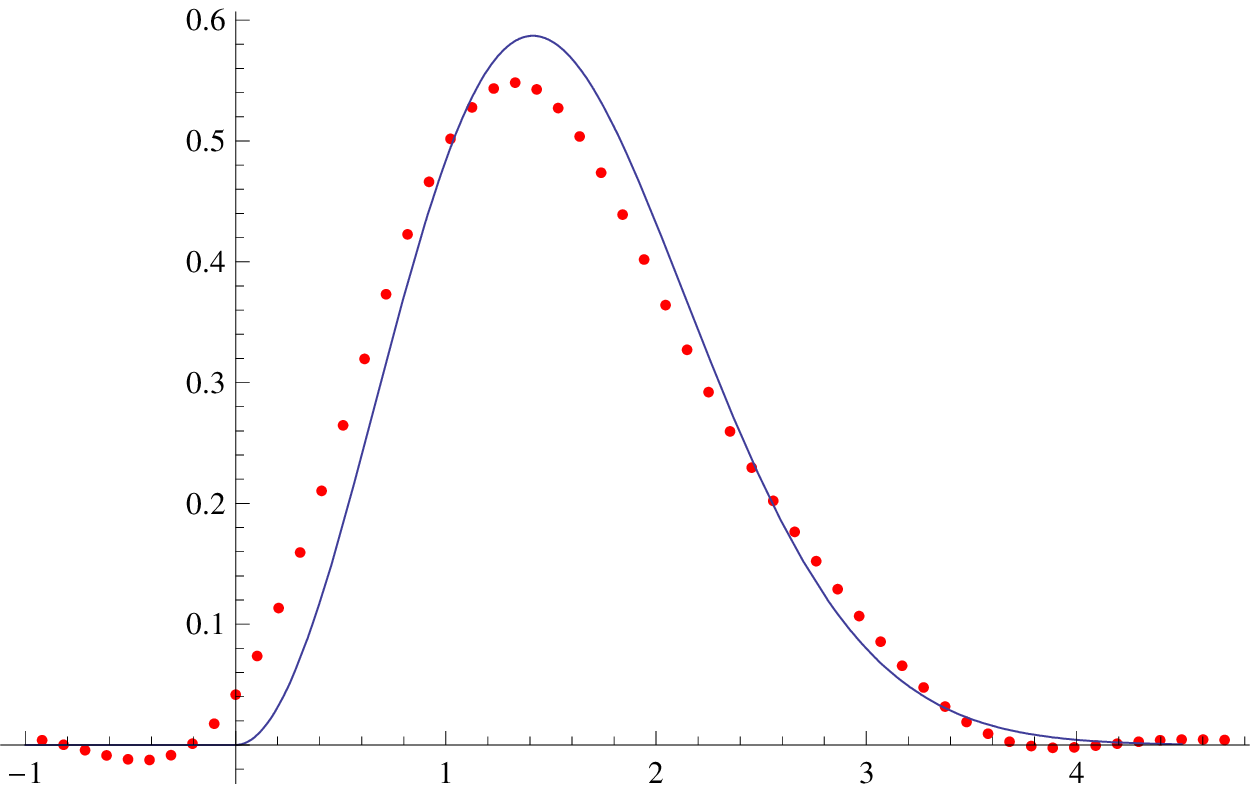}
\end{picture}
\end{minipage}
\vfill
\setlength{\unitlength}{1cm}
\begin{minipage}{5.5cm}
\begin{picture}(5.5,4.0)
\epsfxsize=5.5cm\epsfysize=4cm\epsfbox{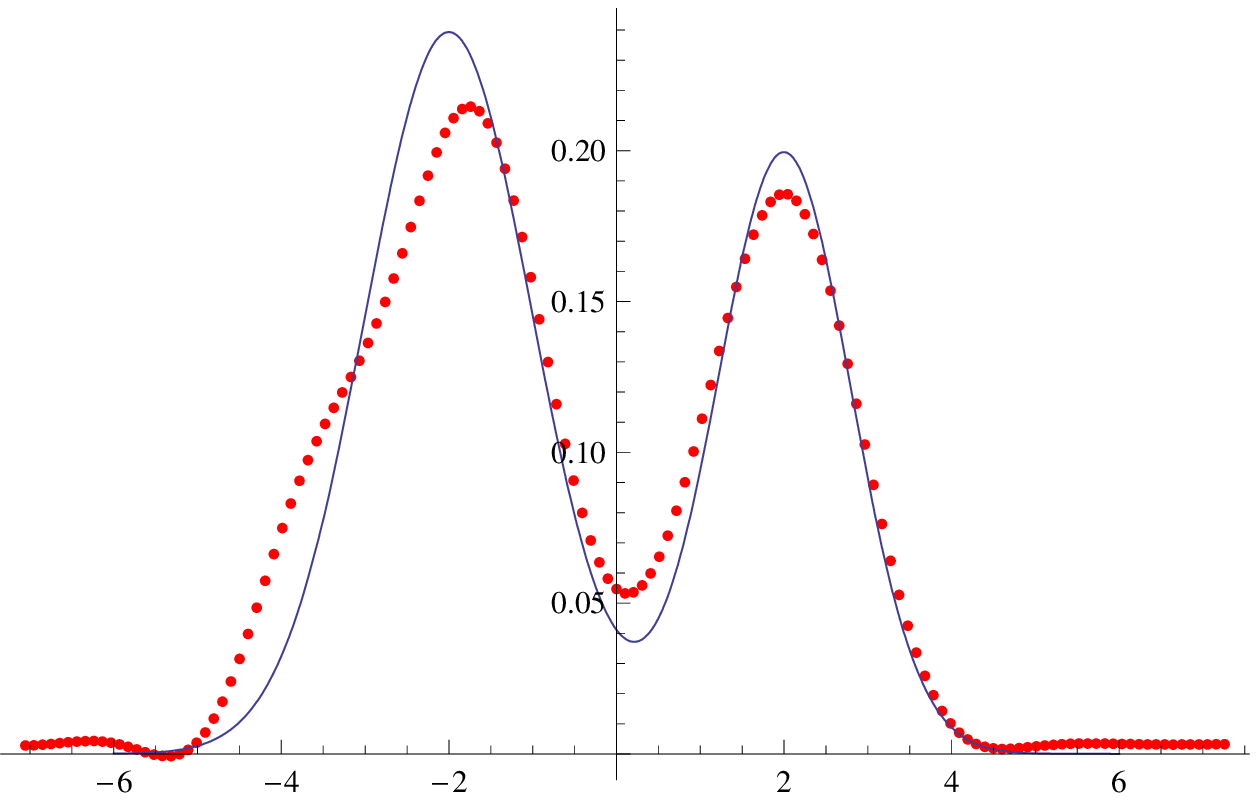}
\end{picture}
\end{minipage}
\hfill
\begin{minipage}{5.5cm}
\begin{picture}(5.5,4.0)
\epsfxsize=5.5cm\epsfysize=4cm\epsfbox{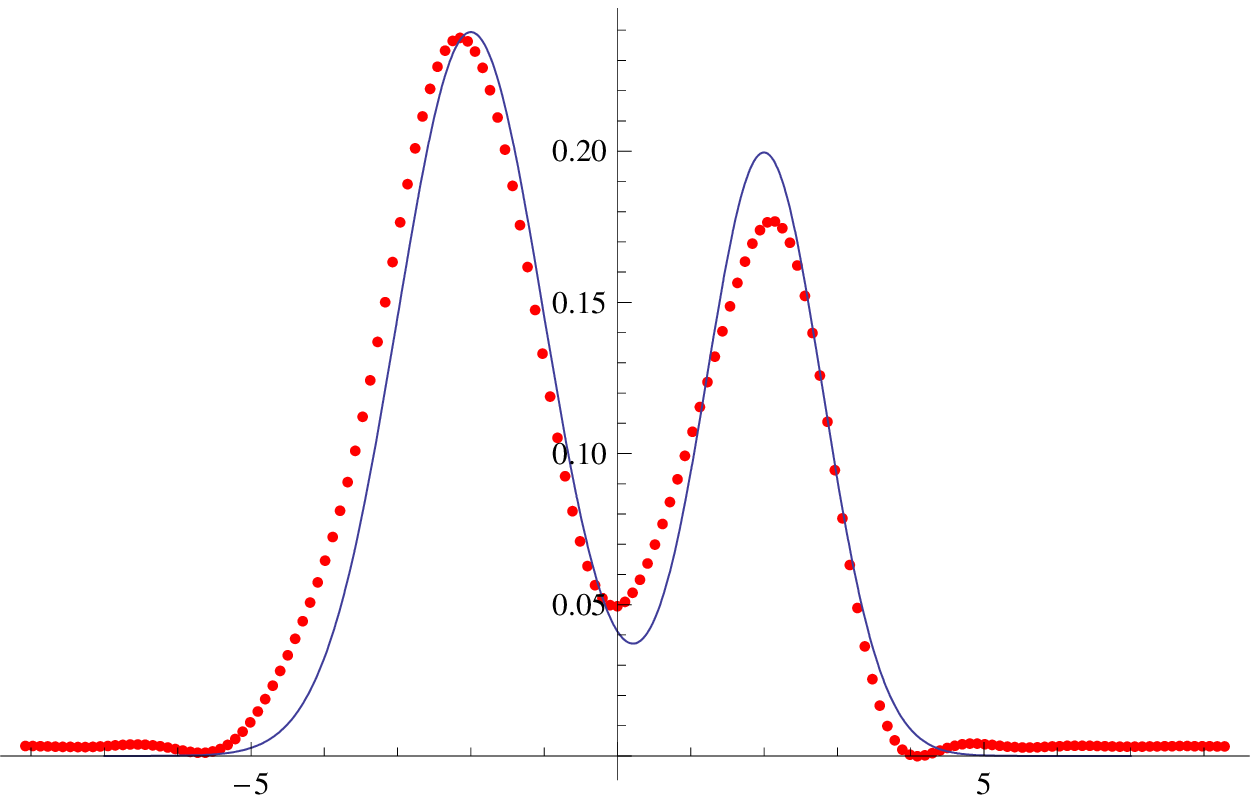}
\end{picture}
\end{minipage}
\caption{\label{fig1} The estimate $f_{nh}$ (dotted line) and the true density $f$ (thin line) for the densities \# 1, \# 2 and \# 3. The left column gives results for $n=50,$ while the right column provides results for $n=100.$}
\end{figure}

Following the recommendation of \citet{delaigle2}, in order to
avoid possible numerical issues, the Fast Fourier Transform was
used to evaluate the estimate \eqref{fnh}. Several outcomes for
two sample sizes, $n=50$ and $n=100,$ are given in Figure
\ref{fig1}. We see that the fit in general is quite reasonable.
This is in line with results in \citet{wand}, where it was shown by finite sample calculations
that the deconvolution kernel density estimator performs well even
in the supersmooth noise distribution case, if the noise level is
not too high.

\begin{figure}[htb]
\setlength{\unitlength}{1cm}
\begin{minipage}{5.5cm}
\begin{picture}(5.5,4.0)
\epsfxsize=5.5cm\epsfysize=4cm\epsfbox{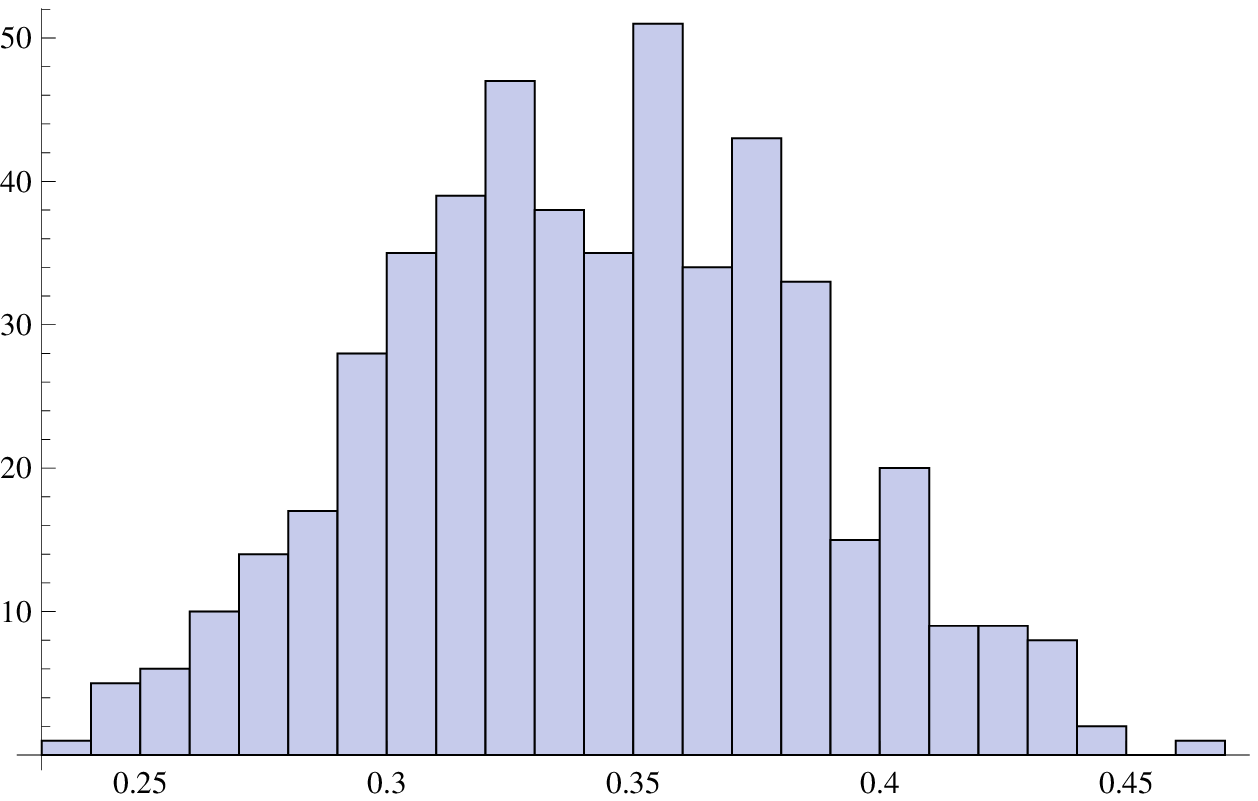}
\end{picture}
\end{minipage}
\hfill
\begin{minipage}{5.5cm}
\begin{picture}(5.5,4.0)
\epsfxsize=5.5cm\epsfysize=4cm\epsfbox{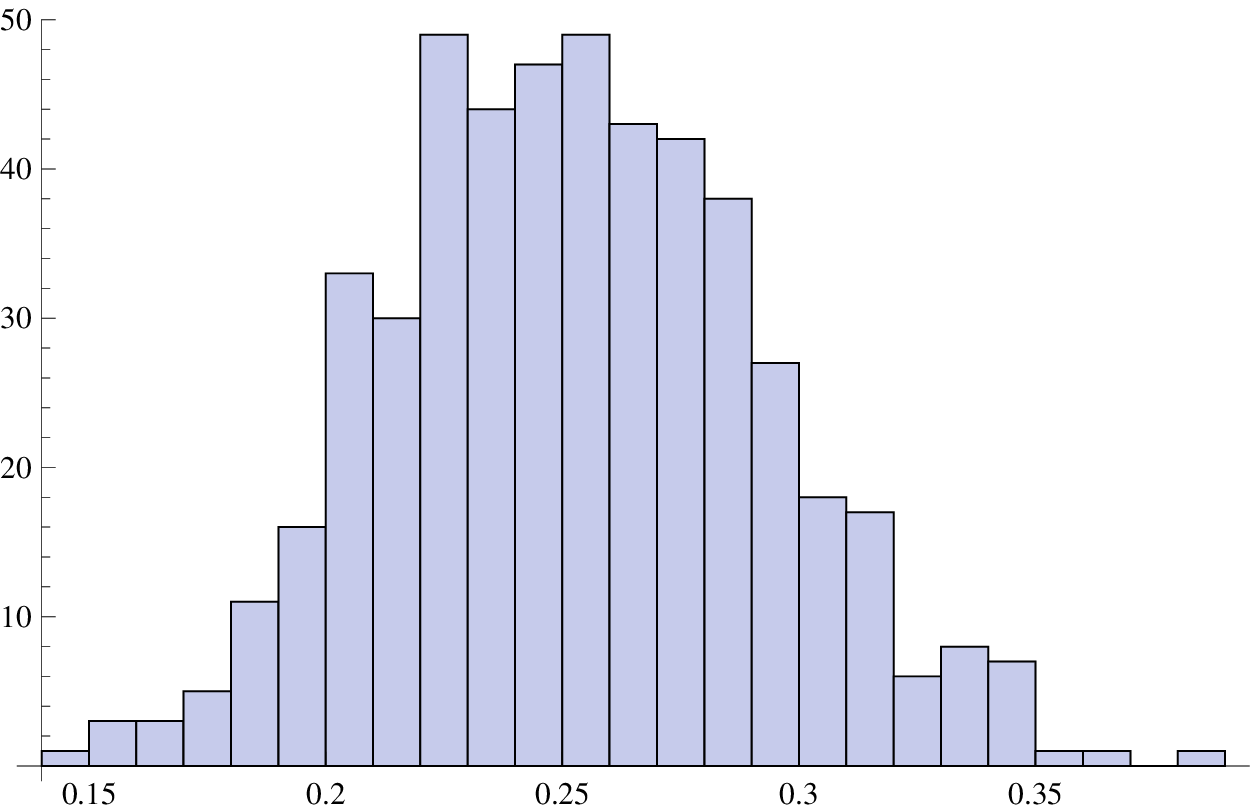}
\end{picture}
\end{minipage}
\vfill
\setlength{\unitlength}{1cm}
\begin{minipage}{5.5cm}
\begin{picture}(5.5,4.0)
\epsfxsize=5.5cm\epsfysize=4cm\epsfbox{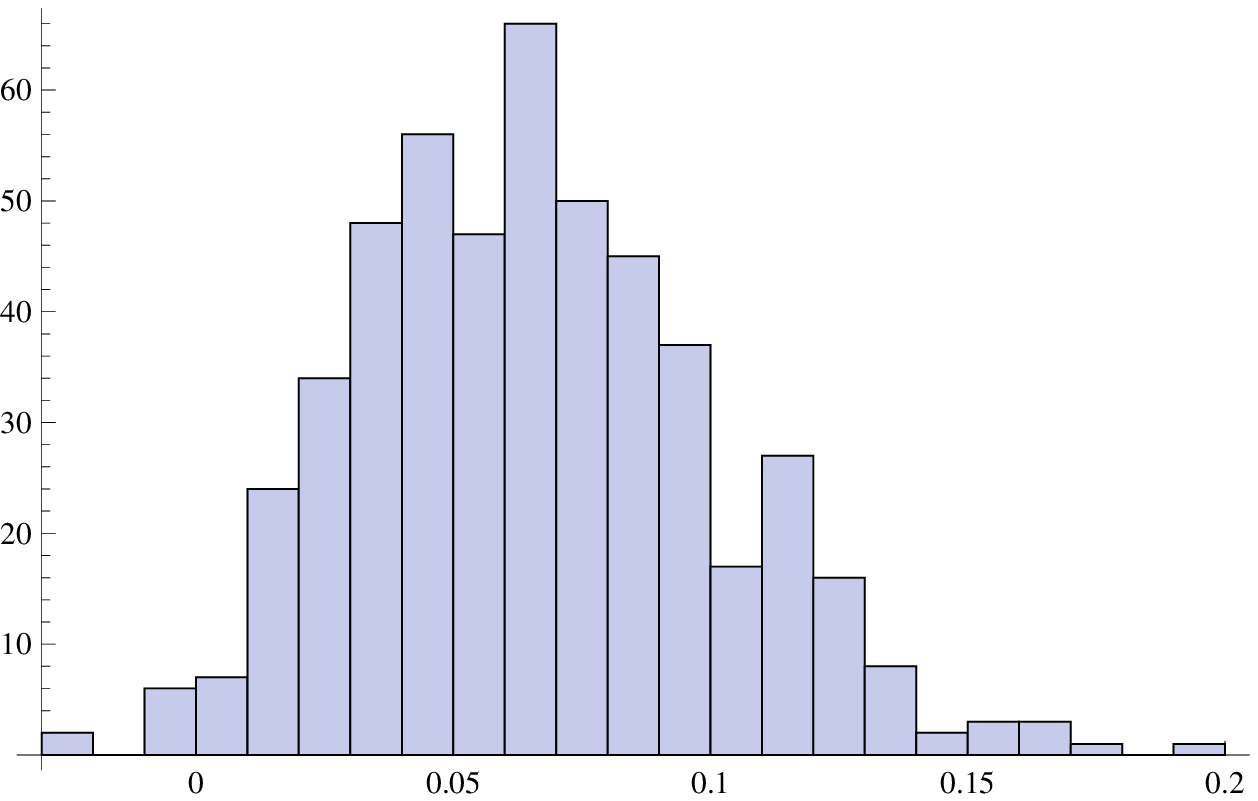}
\end{picture}
\end{minipage}
\hfill
\begin{minipage}{5.5cm}
\begin{picture}(5.5,4.0)
\epsfxsize=5.5cm\epsfysize=4cm\epsfbox{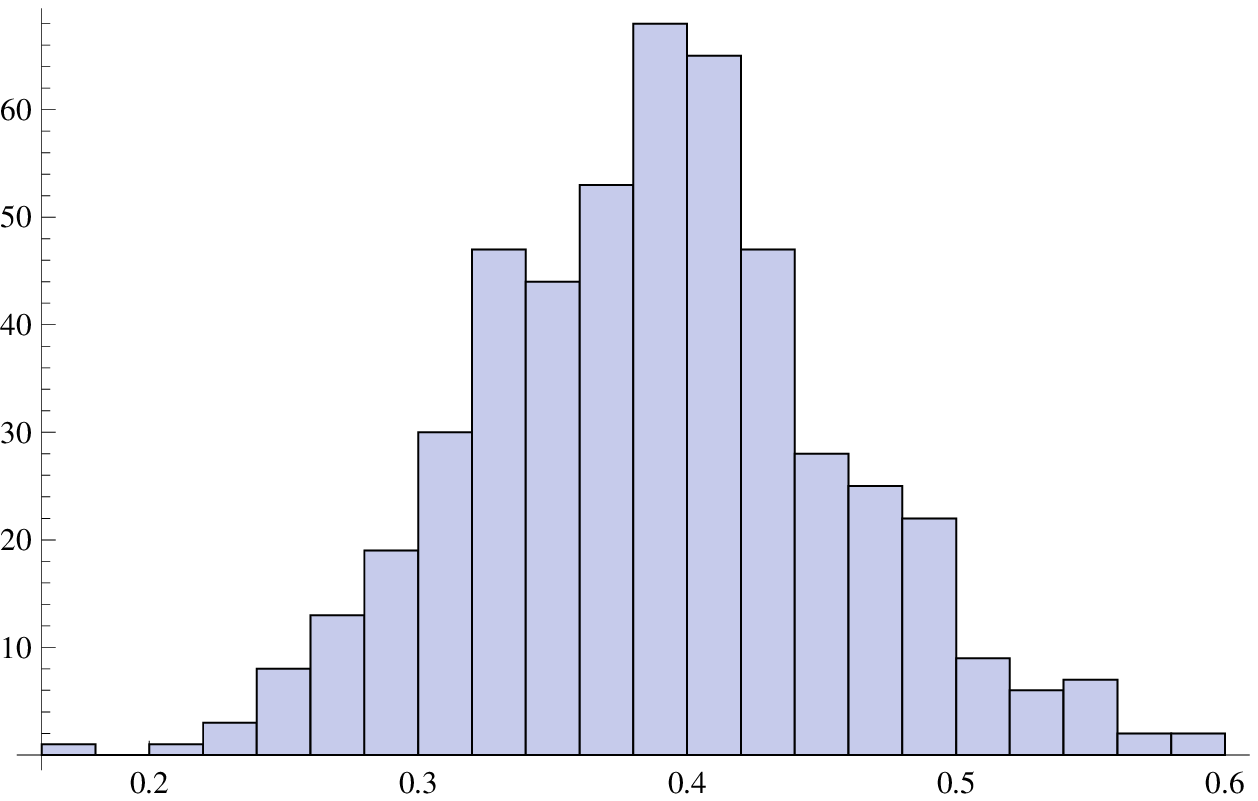}
\end{picture}
\end{minipage}
\vfill
\setlength{\unitlength}{1cm}
\begin{minipage}{5.5cm}
\begin{picture}(5.5,4.0)
\epsfxsize=5.5cm\epsfysize=4cm\epsfbox{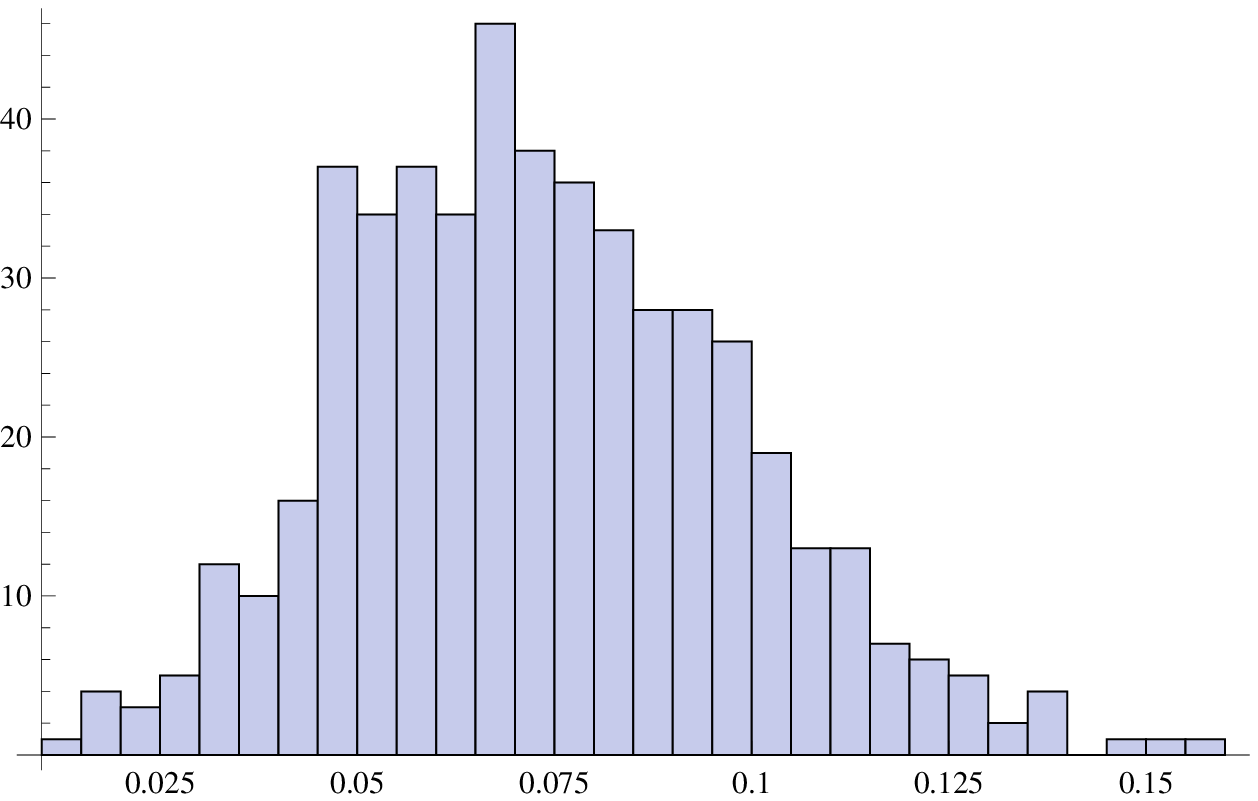}
\end{picture}
\end{minipage}
\hfill
\begin{minipage}{5.5cm}
\begin{picture}(5.5,4.0)
\epsfxsize=5.5cm\epsfysize=4cm\epsfbox{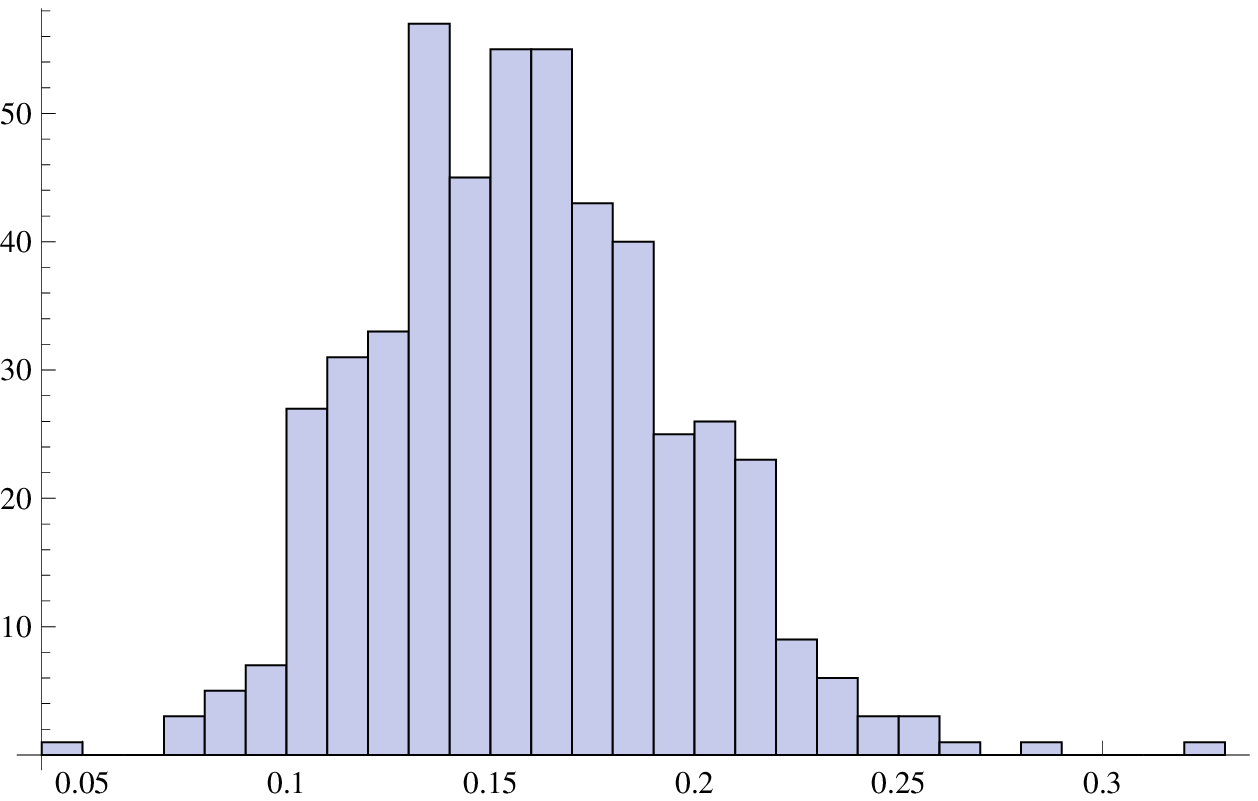}
\end{picture}
\end{minipage}
\caption{\label{fig2} The histograms of estimates $f_{nh}(x)$ for $x=0$ and $x=0.92$ for the density \# 1 (top two graphs), for $x=0$ and $x=0.92$ for the density \# 2 (middle two graphs), and for $x=0$ and $x=2.04$ for the density \# 3 (bottom two graphs).}
\end{figure}

In Figure \ref{fig2} we provide histograms of estimates
$f_{nh}(x)$ that we obtained from our simulations for $x=0$ and
$x=0.92$ (the densities \# 1 and \# 2) and for $x=0$ and $x=2.04$
(the density \# 3). For the density \# 1 points $x=0$ and $x=0.92$
were selected because the first corresponds to its mode, while the
second comes from the region where the value of the density is
moderately high. Notice that $x=0$ is a boundary point for
the support of density \# 2 and that the derivative of density \#
2 is infinite there. For the density \# 3 the point $x=0$
corresponds to the region between its two modes, while $x=2.04$ is
close to where it has one of its modes. The histograms look
satisfactory and indicate that the asymptotic normality is not an
issue.

Our main interest, however, is in comparison of the sample standard deviation of \eqref{fnh} at a fixed point $x$ to the theoretical standard deviation computed using Theorem \ref{thman}. This is of practical importance e.g.\ for construction of confidence intervals. The theoretical standard deviation can be evaluated as
\begin{equation*}
\operatorname{TSD}=\frac{A \Gamma(\alpha+1)
h^{\lambda+\alpha+\lambda_0-1}e^{1/(\mu
h^{\lambda})}}{\sqrt{2n\pi^2}}\left(\frac{\mu}{\lambda}\right)^{1+\alpha},
\end{equation*}
upon noticing that in our case, i.e.\ when using kernel
\eqref{fankernel} and the error distribution ${\mathcal
N}(0,0.4^2),$ we have
$A=8,\alpha=3,\lambda_0=0,\lambda=2,\mu=2/0.4^2.$ After comparing
this theoretical value to the sample standard deviation of the
estimator $f_{nh}$ at points $x=0$ and $x=0.92$ (the densities \#
1 and \# 2) and at points $x=0$ and $x=2.04$ (the density \# 3),
see Table \ref{table1}, we notice a considerable discrepancy (by a
factor $10$ for the density \# 1 and even larger discrepancy for
densities \# 2 and \# 3). At the same time the sample means
evaluated at these two points are close to the true values of the
target density and broadly correspond to the expected theoretical
value $f\ast w_h(x).$ Note here that the bias of $f_{nh}(x)$ is
equal to the bias of an ordinary kernel density estimator based on
a sample from $f,$ see e.g.\ \citet{fan1}.
\begin{table}[htb]
\begin{center}
\begin{tabular}{|c|c|c|c|c|c|c|c|}
\hline
$f$ & $h$ & $\hat{\mu}_1$ & $\hat{\mu}_2$ & $\hat{\sigma}_1$ & $\hat{\sigma}_2$ & $\sigma$ & $\tilde{\sigma}$\\
\hline
\# 1 & 0.24 & 0.343 & 0.252 & 0.0423 & 0.039 & 0.429 & 0.072\\
\hline
\# 2 & 0.18 & 0.066 & 0.389 & 0.035 & 0.067 & 0.169 & 0.114\\
\hline
\# 3 & 0.25 & 0.074 & 0.159 & 0.025 & 0.037 & 0.512 & 0.068\\
\hline
\end{tabular}
\caption{\label{table1} Sample means $\hat{\mu}_1$ and
$\hat{\mu}_2$ and sample standard deviations $\hat{\sigma}_1$ and
$\hat{\sigma}_2$ evaluated at $x=0$ and $x=0.92$ (densities \# 1
and \# 2) and $x=0$ and $x=2.04$ (the density \# 3) together with
the theoretical standard deviation $\sigma$ and the corrected
theoretical standard deviation $\tilde{\sigma}$. The bandwidth is
given by $h.$}
\end{center}
\end{table}

\begin{figure}[htb]
\setlength{\unitlength}{1cm}
\begin{minipage}{5.5cm}
\begin{picture}(5.5,4.0)
\epsfxsize=5.5cm\epsfysize=4cm\epsfbox{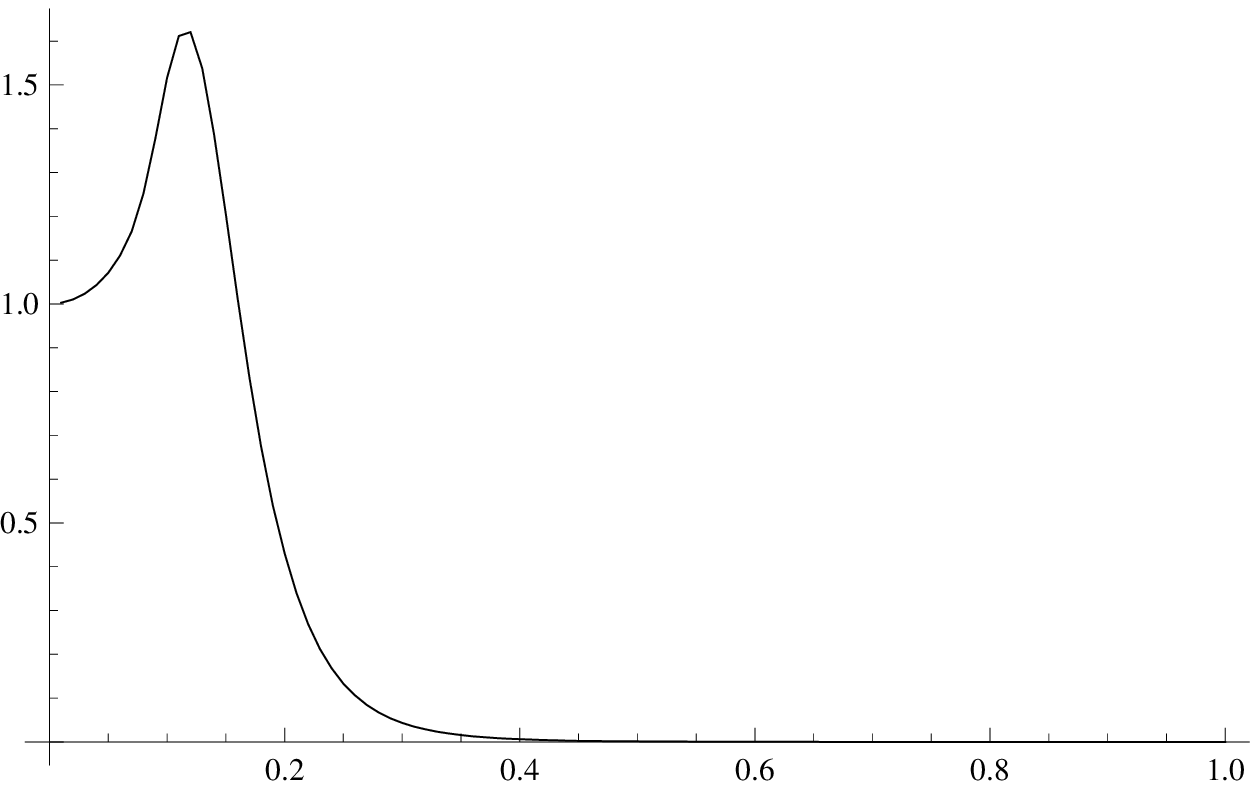}
\end{picture}
\end{minipage}
\hfill
\setlength{\unitlength}{1cm}
\begin{minipage}{5.5cm}
\begin{picture}(5.5,4.0)
\epsfxsize=5.5cm\epsfysize=4cm\epsfbox{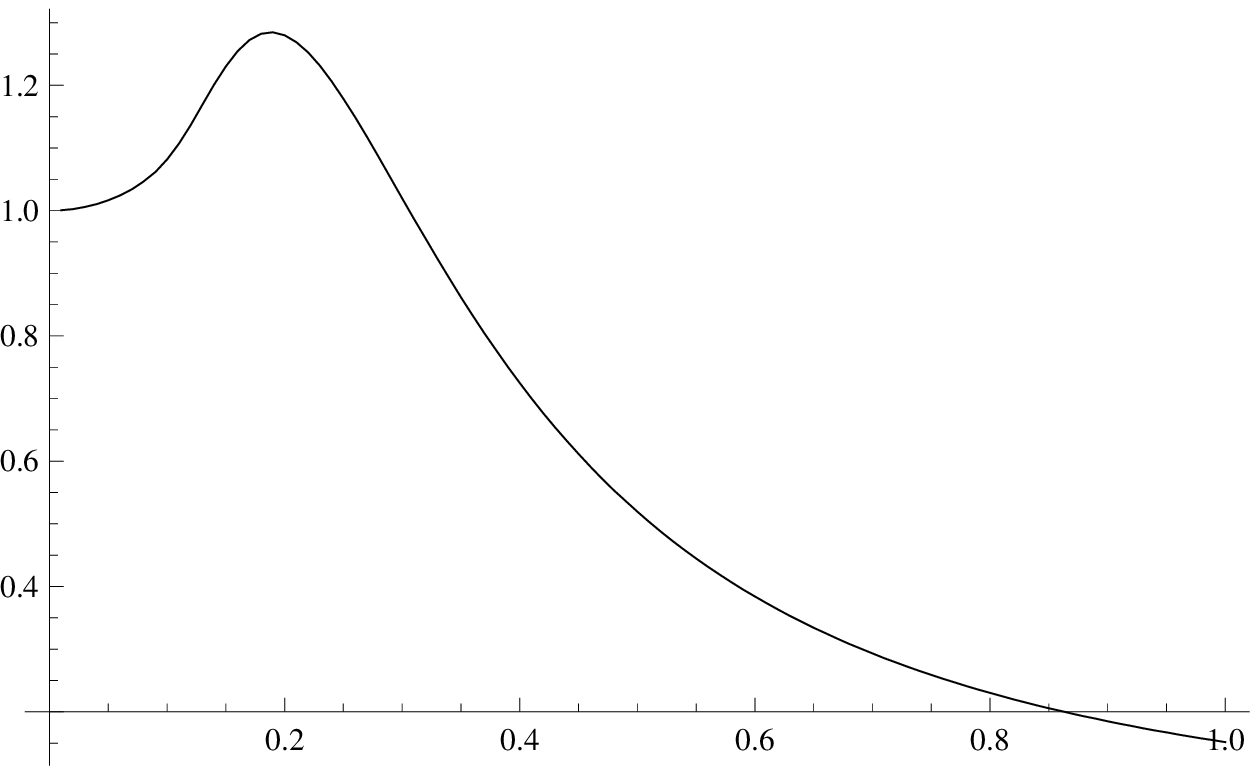}
\end{picture}
\end{minipage}
\caption{\label{ratioplot} Accuracy of \eqref{asnrm2} as a function of $h$ for the kernels \eqref{fankernel} (left figure) and \eqref{sinckernel} (right figure).}
\end{figure}

To gain insight into this striking discrepancy, recall how the
asymptotic normality of $f_{nh}(x)$ was derived in \citet{vanes1}.
Adapting the proof from the latter paper to our example, the first
step is to rewrite $f_{nh}(x)$ as
\begin{equation}
\label{asnrm1}
\frac{1}{\pi h}\int_{0}^1 \phi_w(s) \exp[s^{\lambda}/(\mu h^{\lambda})]ds\frac{1}{n}\sum_{j=1}^n \cos\left(\frac{x-X_j}{h}\right)+\frac{1}{n}\sum_{j}^n \tilde{R}_{n,j},
\end{equation}
where the remainder terms $\tilde{R}_{n,j}$ are defined in \citet{vanes1}. Then by estimating the variance of the second summand in \eqref{asnrm1}, one can show that it can be neglected when considering the asymptotic normality of \eqref{asnrm1} as $n\rightarrow\infty$ and $h\rightarrow 0.$ Turning to the first term in \eqref{asnrm1}, one uses the asymptotic equivalence, cf.\ Lemma 5 in \citet{vanes1},
\begin{equation}
\label{asnrm2}
\int_{0}^1 \phi_w(s) \exp[s^{\lambda}/(\mu h^{\lambda})]ds \sim A \Gamma (\alpha+1) \left(\frac{\mu}{\lambda}h^{\lambda}\right)^{1+\alpha} e^{1/(\mu h^{\lambda})},
\end{equation}
which explains the shape of the normalising constant in Theorem \ref{thman}. However, this is precisely the point which causes a large discrepancy between the theoretical standard deviation and the sample standard deviation. The approximation is good asymptotically as $h\rightarrow0,$ but it is quite inaccurate for larger values of $h.$ Indeed, consider the ratio of the left-hand side of \eqref{asnrm2} with the right-hand side. We have plotted this ratio as a function of $h$ for $h$ ranging between $0$ and $1,$ see Figure \ref{ratioplot}. One sees that the ratio is close to $1$ for extremely small values of $h$ and is quite far from $1$ for larger values of $h.$ It is equally easy to see that the poor approximation in \eqref{asnrm2} holds true for kernels \eqref{sinckernel} and \eqref{wandkernel} as well, see e.g.\ Figure \ref{ratioplot}, which plots the ratio of both sides of \eqref{asnrm2} for the kernel \eqref{sinckernel}.
This poor approximation, of course, is not characteristic of only
the particular $\mu$ and $\lambda$ that we used in our
simulations, but also holds true for other values of $\mu$ and
$\lambda.$

\begin{figure}[htb]
\setlength{\unitlength}{1cm}
\begin{minipage}{5.5cm}
\begin{picture}(5.5,4.0)
\epsfxsize=5.5cm\epsfysize=4cm\epsfbox{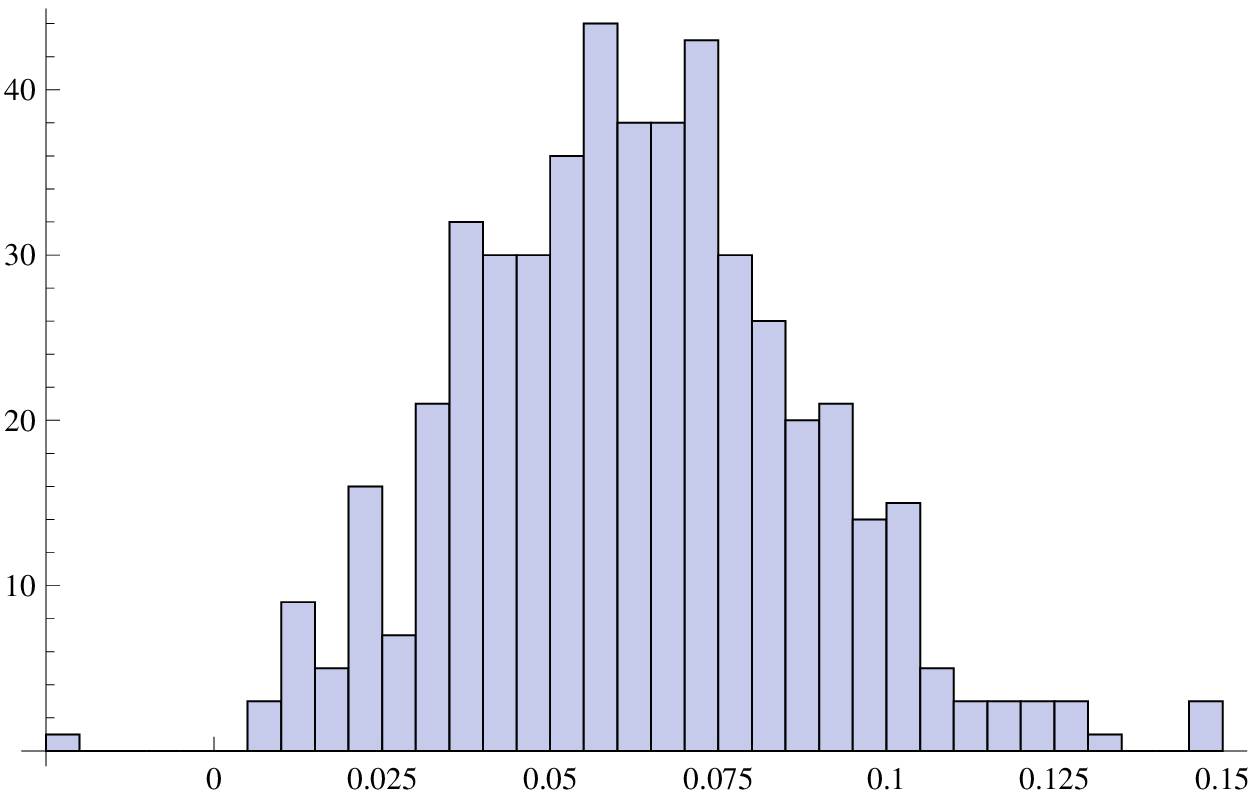}
\end{picture}
\end{minipage}
\hfill
\begin{minipage}{5.5cm}
\begin{picture}(5.5,4.0)
\epsfxsize=5.5cm\epsfysize=4cm\epsfbox{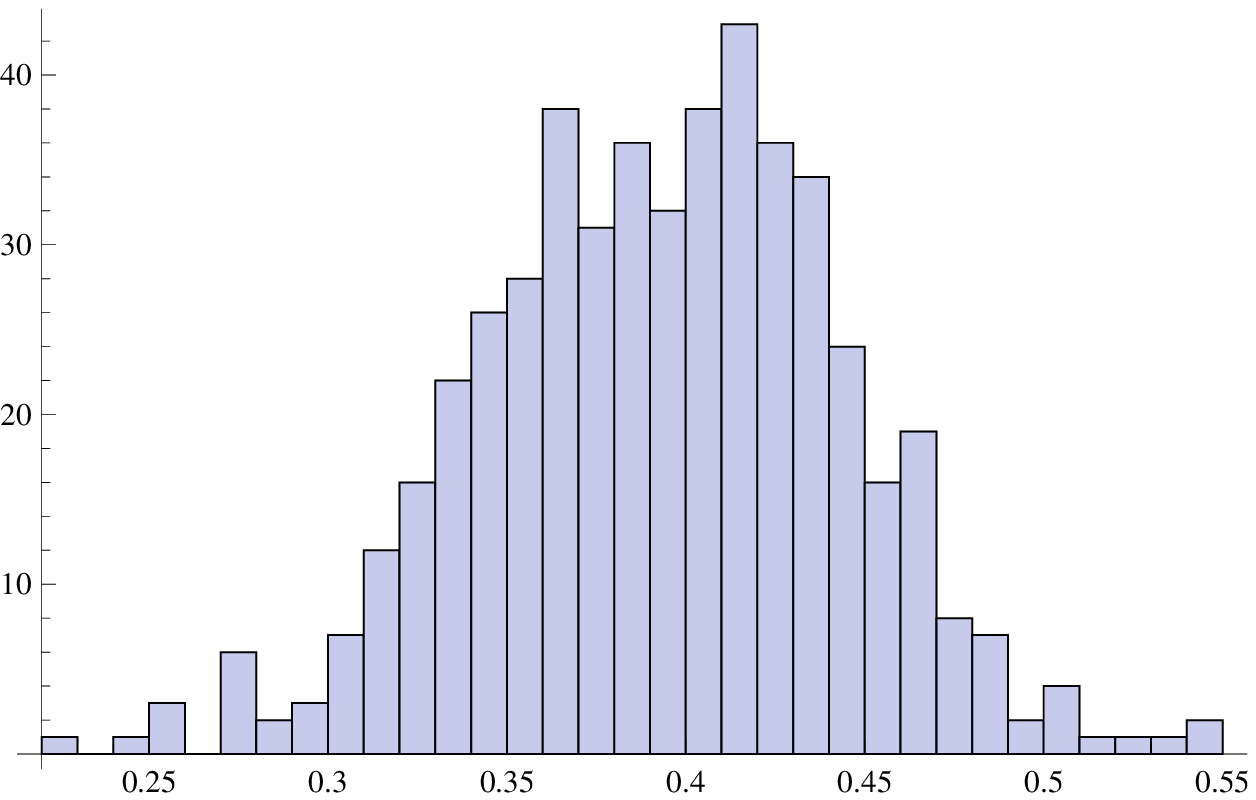}
\end{picture}
\end{minipage}
\vfill
\setlength{\unitlength}{1cm}
\begin{minipage}{5.5cm}
\begin{picture}(5.5,4.0)
\epsfxsize=5.5cm\epsfysize=4cm\epsfbox{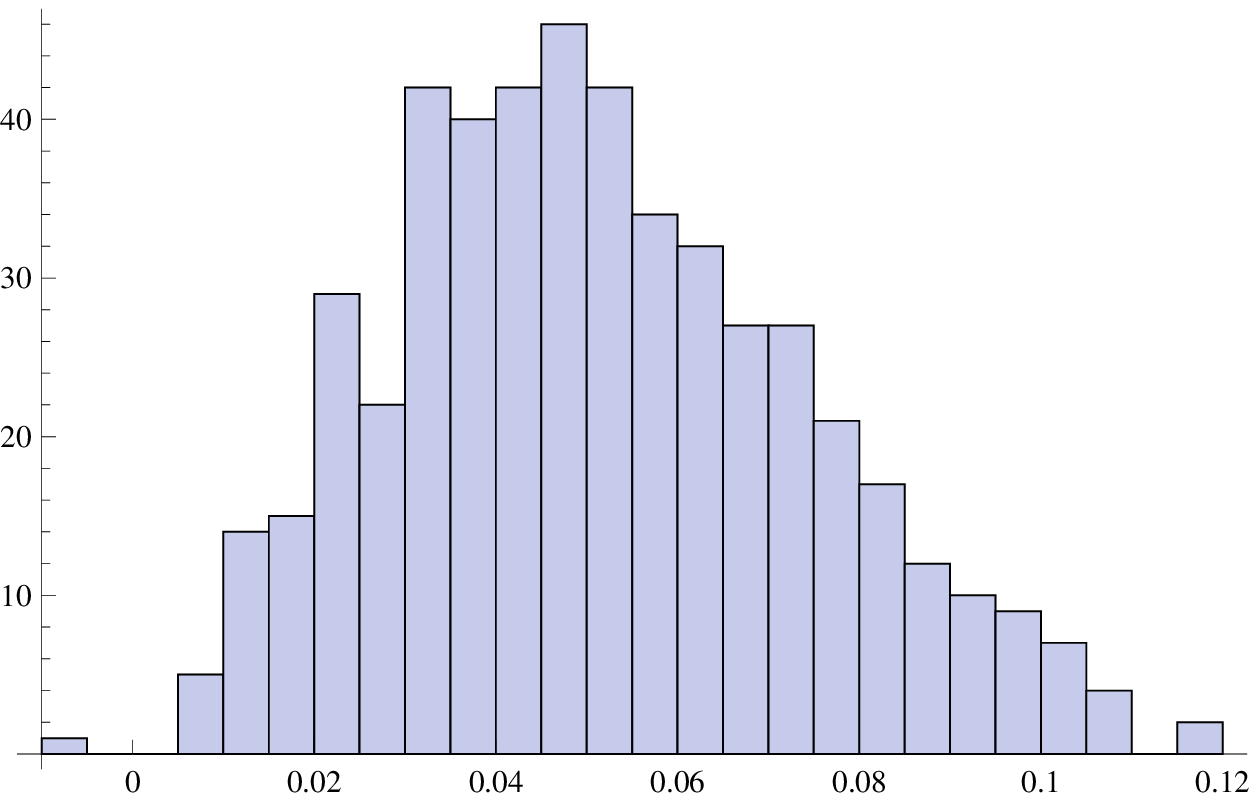}
\end{picture}
\end{minipage}
\hfill
\begin{minipage}{5.5cm}
\begin{picture}(5.5,4.0)
\epsfxsize=5.5cm\epsfysize=4cm\epsfbox{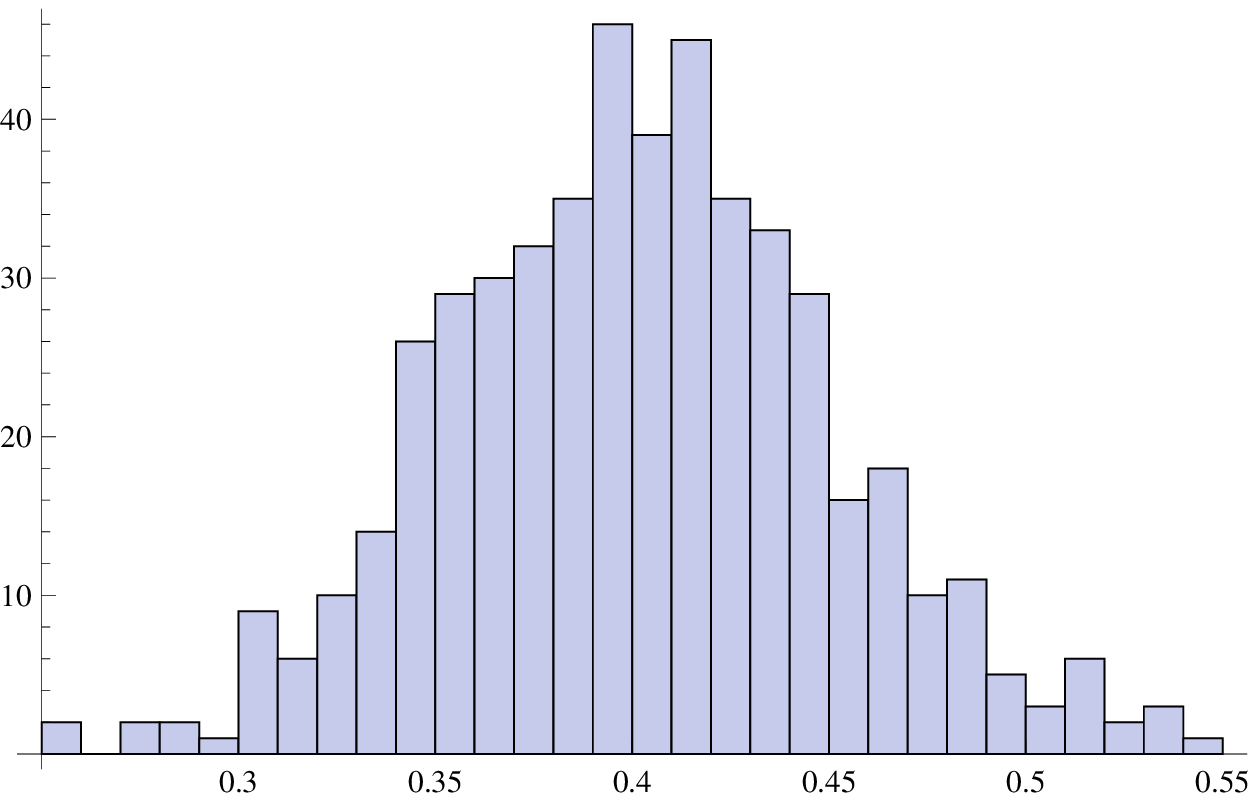}
\end{picture}
\end{minipage}
\caption{\label{chinlarge} The histograms of estimates $f_{nh}(x)$ for $x=0$ and $x=0.92$ for the density \# 2 for $n=100$ (top two graphs) and for $n=200$ (bottom two graphs).}
\end{figure}

Obviously, one can correct for the poor approximation of the
sample standard deviation by the theoretical standard deviation by
using the left-hand side of \eqref{asnrm2} instead of its
approximation. The theoretical standard deviation corrected in
such a way is given in the last column of Table \ref{table1}. As
it can be seen from the table, this procedure led to an
improvement of the agreement between the theoretical standard
deviation and its sample counterpart for all three target
densities. Nevertheless, the match is not entirely satisfactory,
since the corrected theoretical standard deviation and the sample
standard deviation differ by factor $2$ or even more. A perfect
match is impossible to obtain, because we neglect the remainder
term in \eqref{asnrm1} and $h$ is still fairly large. We further
notice that the concurrence between the results is better for
$x=0$ than for $x=0.92$ for densities \# 1 and \# 2, and for
$x=2.04$ than for $x=0$ for the density \# 3. We also performed
simulations for the sample sizes $n=100$ and $n=200$ to check the
effect of having larger samples. For brevity we will report only
the results for density \# 2, see Figure \ref{chinlarge} and Table
\ref{chitable}, since this density is nontrivial to deconvolve,
though not as difficult as the density \# 3. Notice that the
results did not improve greatly for $n=100,$ while for the case
$n=200$ the corrected theoretical standard deviation became a
worse estimate of the sample standard deviation than the
theoretical standard deviation. Explanation of this curious
phenomenon is given in Section \ref{conclusions}.

\begin{table}[htb]
\begin{center}
\begin{tabular}{|c|c|c|c|c|c|c|c|}
\hline
$n$ & $h$ & $\hat{\mu}_1$ & $\hat{\mu}_2$ & $\hat{\sigma}_1$ & $\hat{\sigma}_2$ & $\sigma$ & $\tilde{\sigma}$\\
\hline
\# 100 & 0.17 & 0.063 & 0.393 & 0.025 & 0.051 & 0.108 & 0.090\\
\hline
\# 200 & 0.15 & 0.052 & 0.402 & 0.023 & 0.049 & 0.070 & 0.084\\
\hline
\end{tabular}
\caption{\label{chitable} Sample means $\hat{\mu}_1$ and $\hat{\mu}_2$ and sample standard deviations $\hat{\sigma}_1$ and $\hat{\sigma}_2$ evaluated at $x=0$ and $x=0.92$ for the density \# 2, together with the theoretical standard deviation $\sigma$ and the corrected theoretical standard deviation $\tilde{\sigma}$.}
\end{center}
\end{table}

Furthermore, note that
\begin{equation*}
\var\left[\frac{1}{\sqrt{n}}\sum_{j=1}^n \cos\left(\frac{x-X_j}{h}\right)\right]\rightarrow \frac{1}{2}
\end{equation*}
as $n\rightarrow \infty$ and $h\rightarrow0,$ see \citet{vanes1}. This explains the appearance of the factor $1/2$ in the asymptotic variance in Theorem \ref{thman}. One might also question the goodness of this approximation and propose to use instead some estimator of $\var[\cos((x-X)h^{-1})],$ e.g.\ its empirical counterpart based on the sample $X_1,\ldots,X_n.$ However, in the simulations that we performed for all three target densities (with $n$ and $h$ as above), the resulting estimates took values close to the true value $1/2.$ E.g.\ for the density \# 3 the sample mean turned out to be $0.502298,$ while the sample standard deviation was equal to $0.0535049,$ thus showing that there was insignificant variability around $1/2$ in this particular example. On the other hand, for other distributions and for different sample sizes, it could be the case that the direct use of $1/2$ will lead to inaccurate results.

\begin{figure}[htb]
\setlength{\unitlength}{1cm}
\begin{minipage}{5.5cm}
\begin{picture}(5.5,4.0)
\epsfxsize=5.5cm\epsfysize=4cm\epsfbox{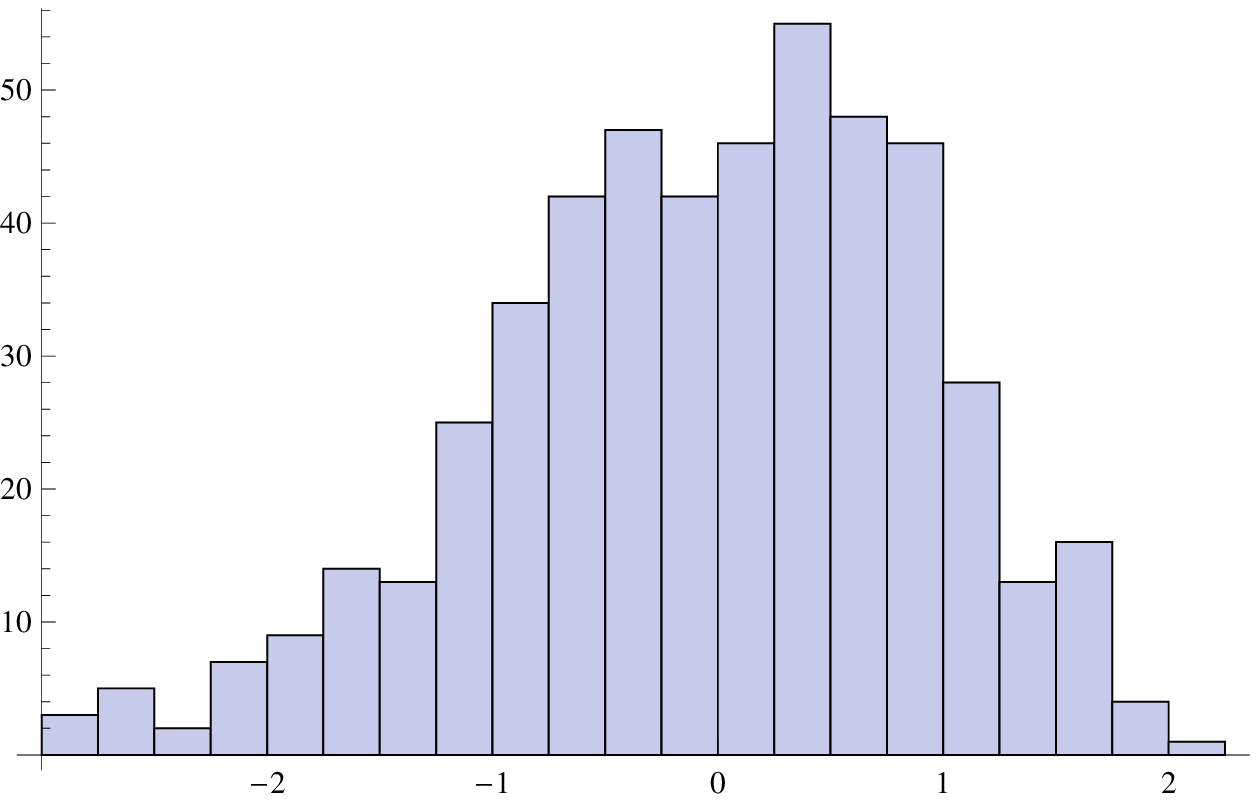}
\end{picture}
\end{minipage}
\hfill
\begin{minipage}{5.5cm}
\begin{picture}(5.5,4.0)
\epsfxsize=5.5cm\epsfysize=4cm\epsfbox{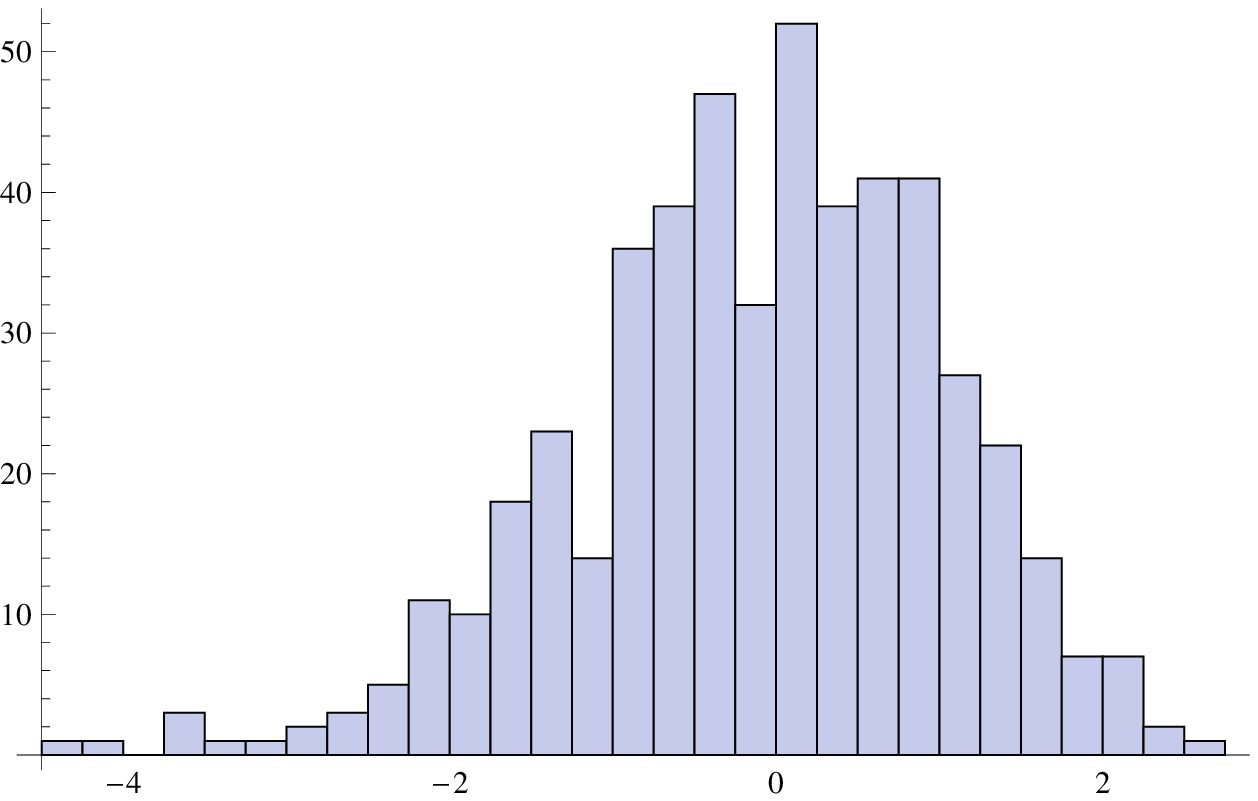}
\end{picture}
\end{minipage}
\vfill
\setlength{\unitlength}{1cm}
\begin{minipage}{5.5cm}
\begin{picture}(5.5,4.0)
\epsfxsize=5.5cm\epsfysize=4cm\epsfbox{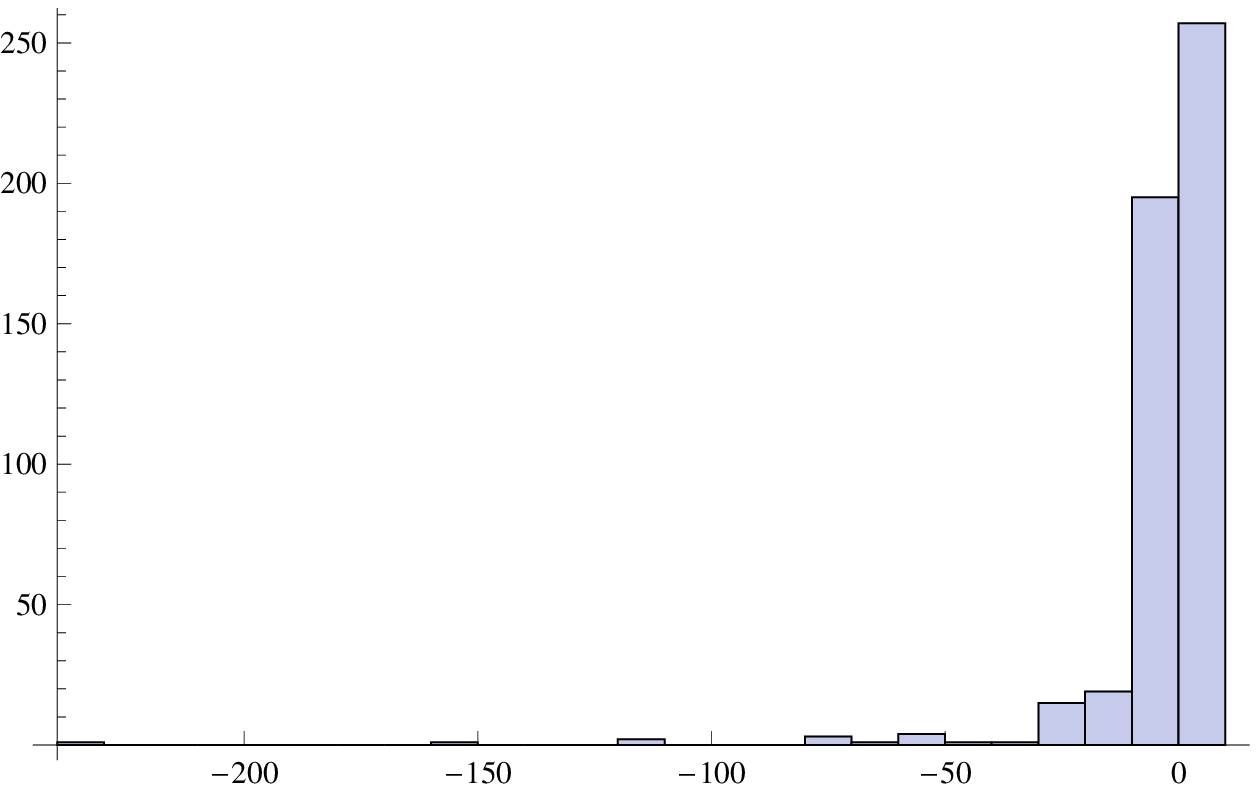}
\end{picture}
\end{minipage}
\hfill
\begin{minipage}{5.5cm}
\begin{picture}(5.5,4.0)
\epsfxsize=5.5cm\epsfysize=4cm\epsfbox{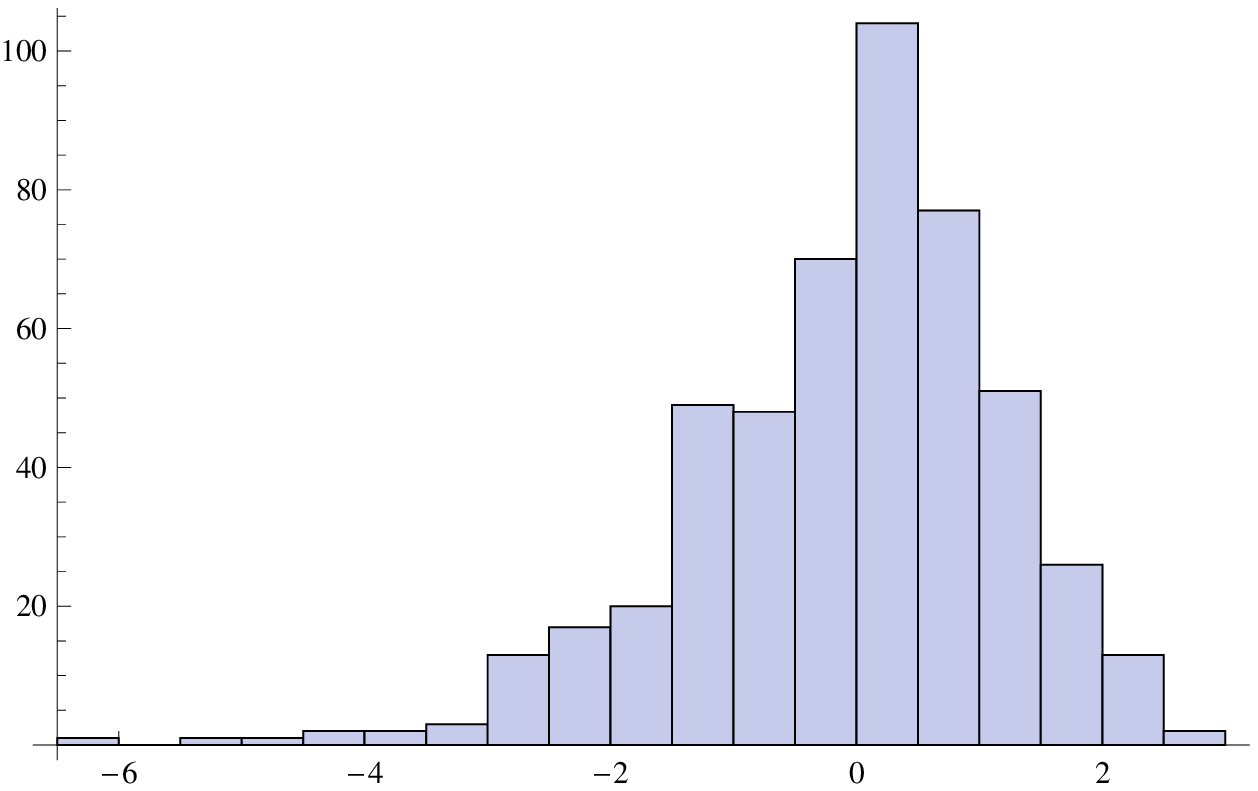}
\end{picture}
\end{minipage}
\vfill
\setlength{\unitlength}{1cm}
\begin{minipage}{5.5cm}
\begin{picture}(5.5,4.0)
\epsfxsize=5.5cm\epsfysize=4cm\epsfbox{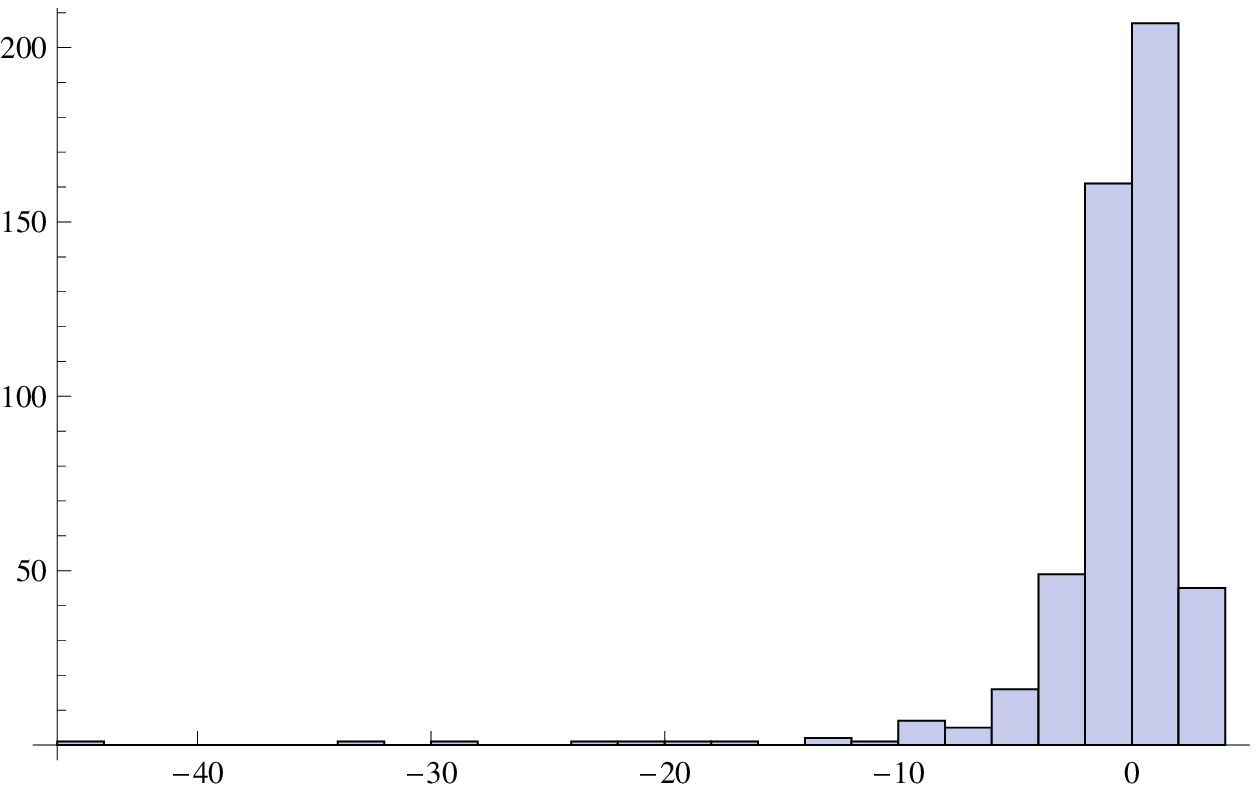}
\end{picture}
\end{minipage}
\hfill
\begin{minipage}{5.5cm}
\begin{picture}(5.5,4.0)
\epsfxsize=5.5cm\epsfysize=4cm\epsfbox{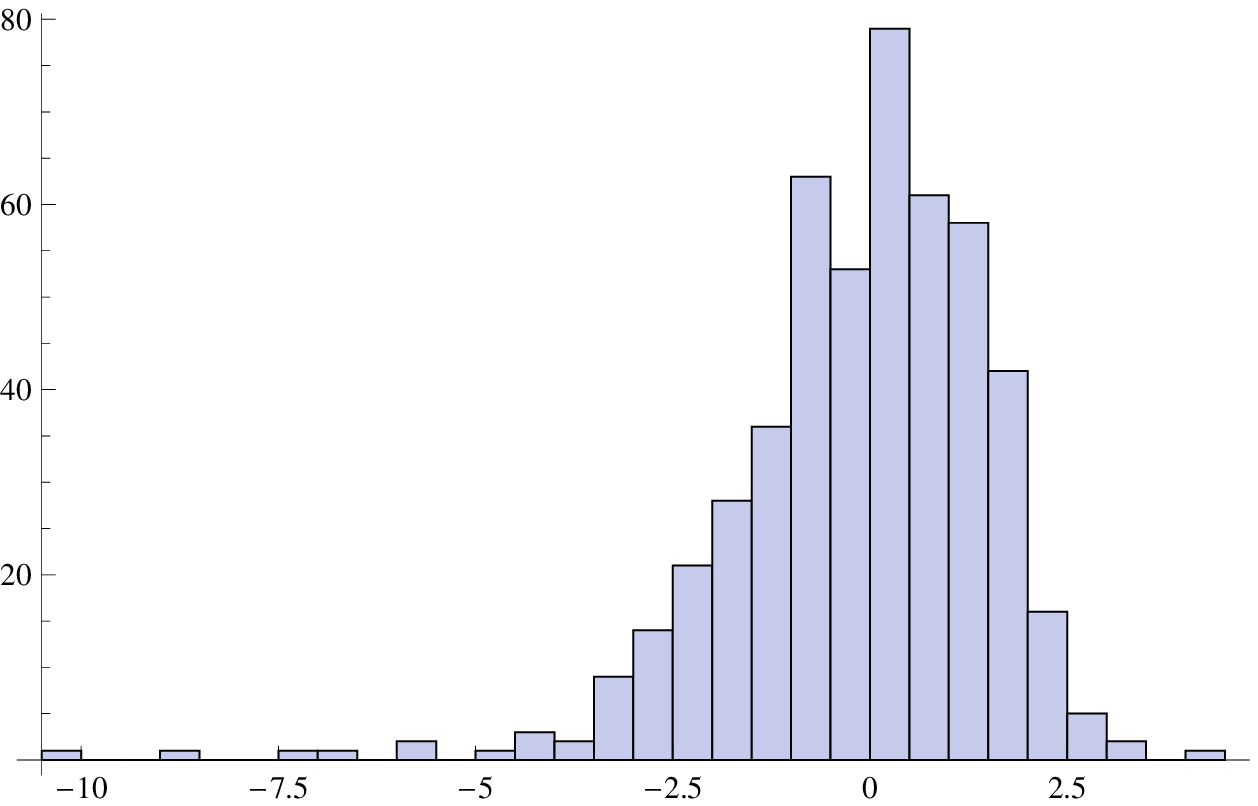}
\end{picture}
\end{minipage}
\caption{\label{fanhisto} The histograms of \eqref{fanexpr} for $x=0$ and $x=0.92$ for the density \# 1 (top two graphs), for $x=0$ and $x=0.92$ for the density \# 2 (middle two graphs), and for $x=0$ and $x=2.04$ for the density \# 3 (bottom two graphs).}
\end{figure}

\begin{figure}[htb]
\setlength{\unitlength}{1cm}
\begin{minipage}{5.5cm}
\begin{picture}(5.5,4.0)
\epsfxsize=5.5cm\epsfysize=4cm\epsfbox{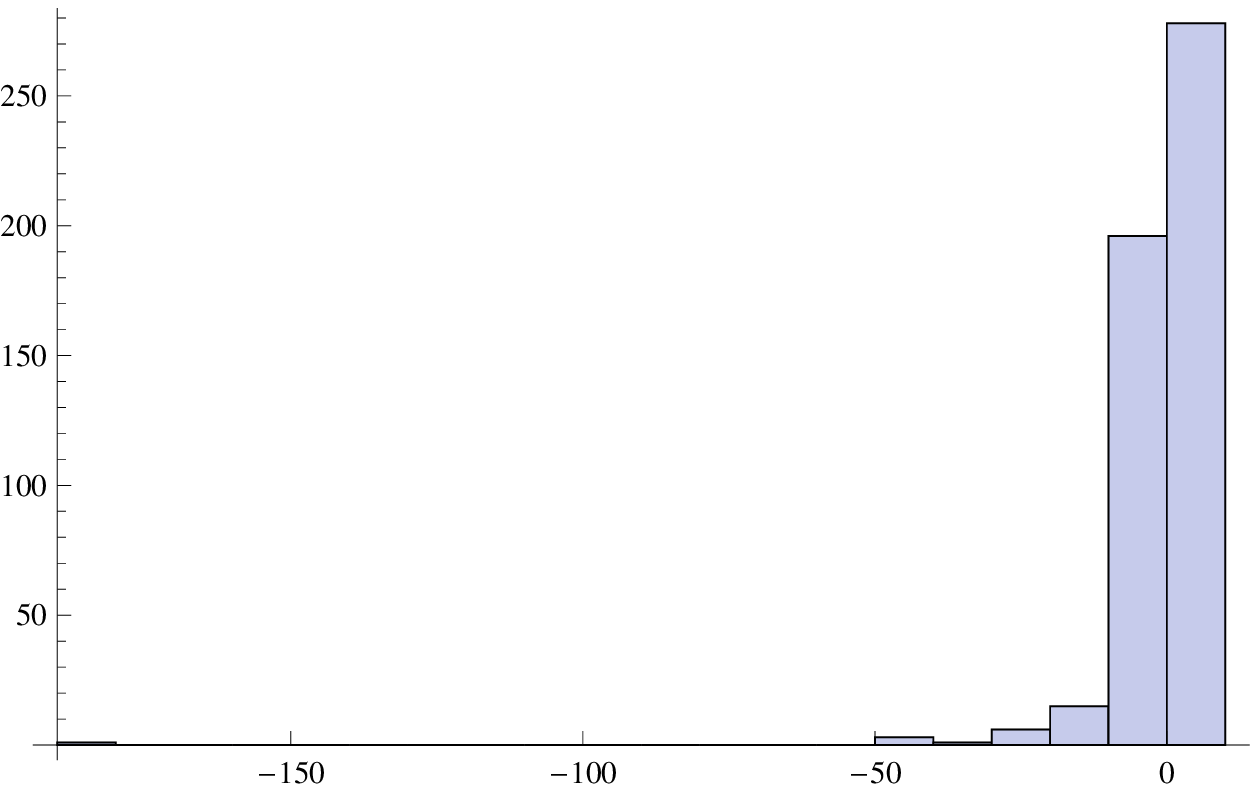}
\end{picture}
\end{minipage}
\hfill
\begin{minipage}{5.5cm}
\begin{picture}(5.5,4.0)
\epsfxsize=5.5cm\epsfysize=4cm\epsfbox{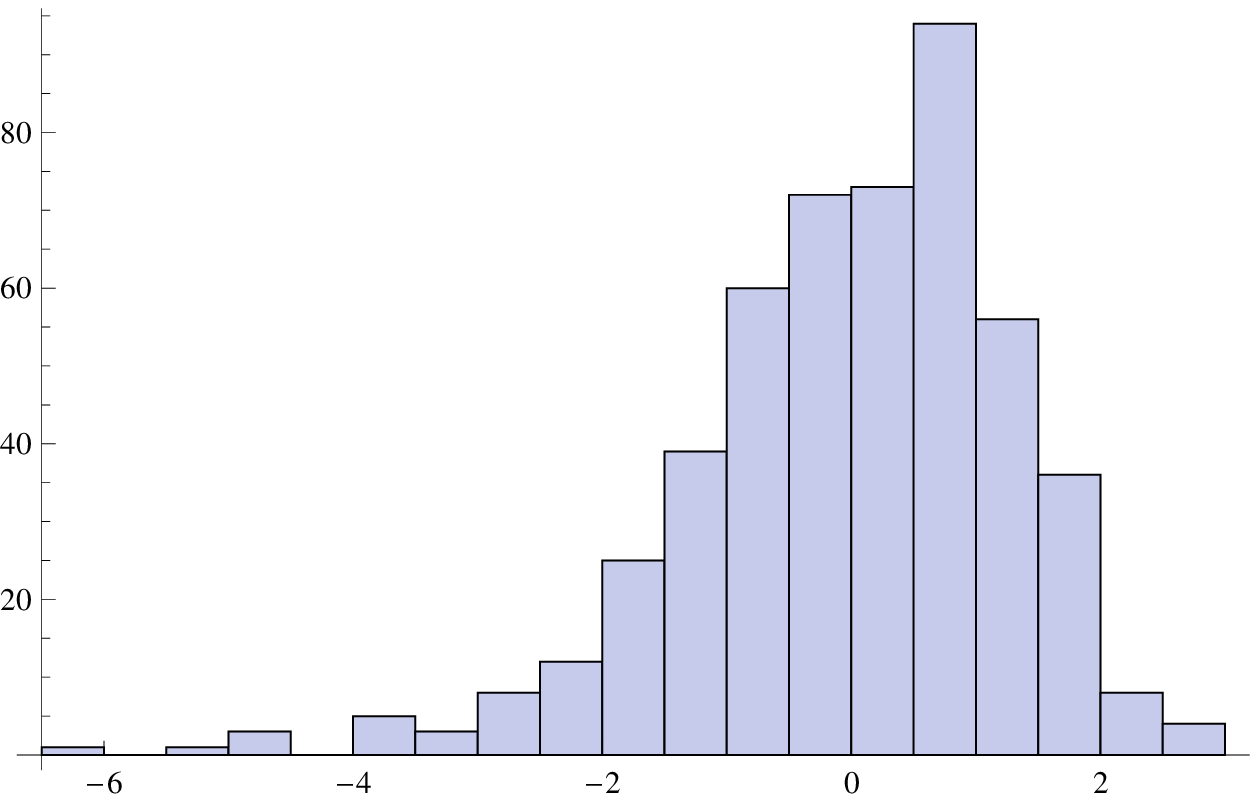}
\end{picture}
\end{minipage}
\vfill
\setlength{\unitlength}{1cm}
\begin{minipage}{5.5cm}
\begin{picture}(5.5,4.0)
\epsfxsize=5.5cm\epsfysize=4cm\epsfbox{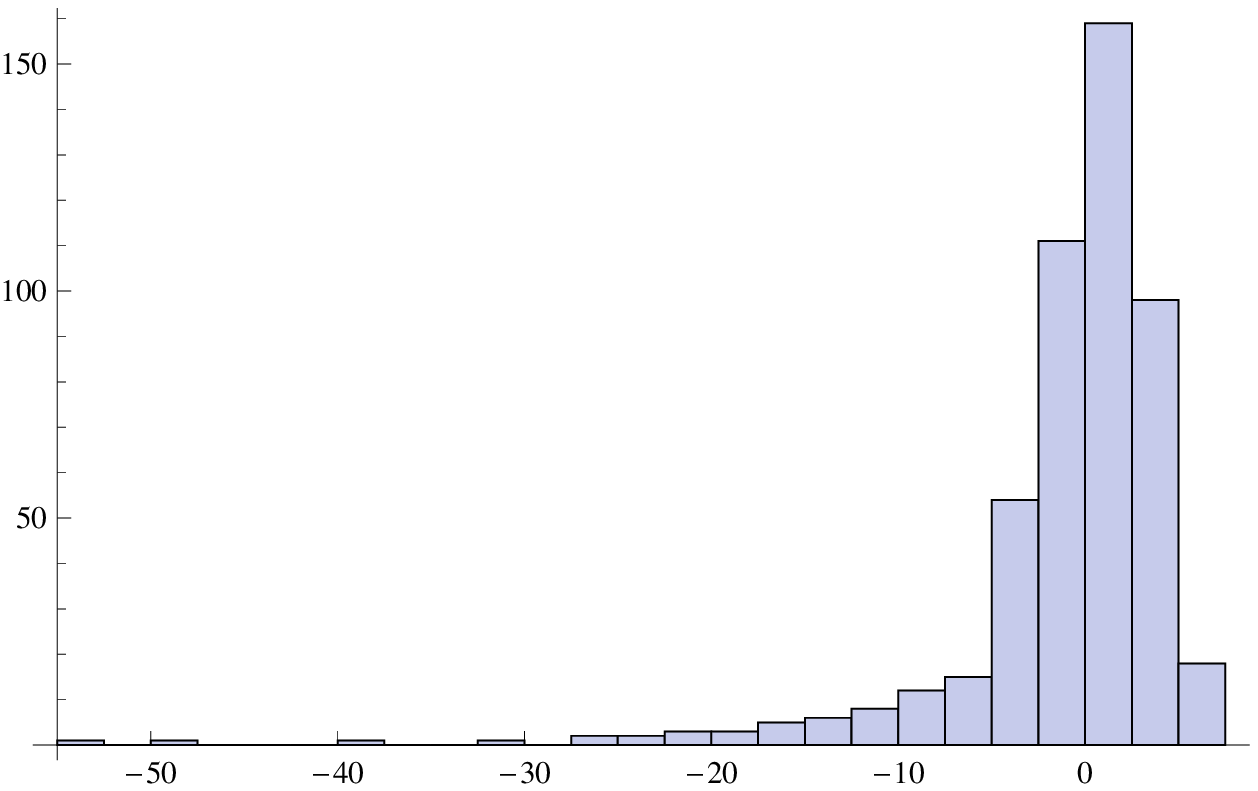}
\end{picture}
\end{minipage}
\hfill
\begin{minipage}{5.5cm}
\begin{picture}(5.5,4.0)
\epsfxsize=5.5cm\epsfysize=4cm\epsfbox{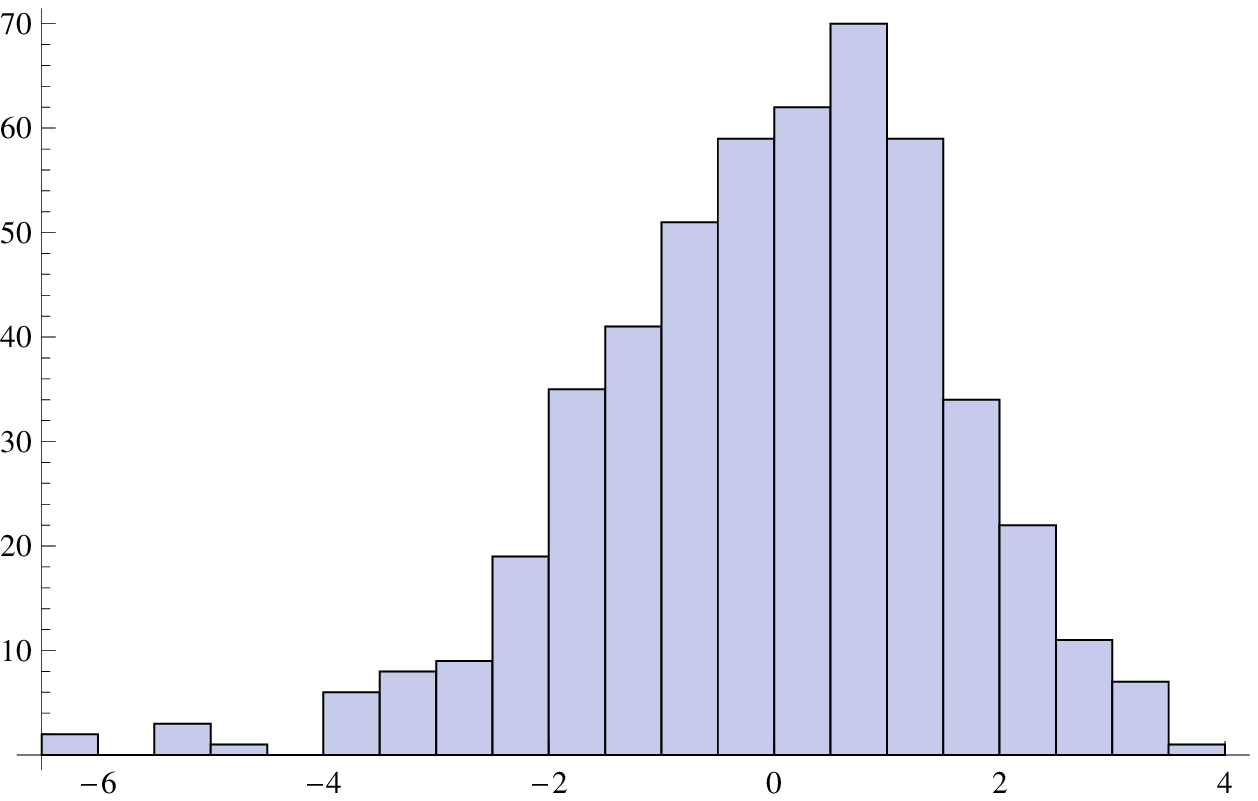}
\end{picture}
\end{minipage}
\caption{\label{chifan} The histograms of \eqref{fanexpr} for $x=0$ and $x=0.92$ for the density \# 2 for $n=100$ (top two graphs)and $n=200$ (bottom two graphs).}
\end{figure}

Next we report some simulation results relevant to Theorem
\ref{thmanfan}. This theorem tells us that for a fixed $n$ we have
that
\begin{equation}
\label{fanexpr}
\frac{\sqrt{n}}{s_n}(f_{nh}(x)-\ex[f_{nh}(x)])
\end{equation}
is approximately normally distributed with zero mean and variance
equal to one. Upon using the fact that $\ex[f_{nh}(x)]=f\ast
w_h(x),$ we used the data that we obtained from our previous
simulation examples to plot the histograms of \eqref{fanexpr} and
to evaluate the sample means and standard deviations, see Figure
\ref{fanhisto} and Table \ref{fantable}. One notices that the
concurrence of the theoretical and sample values is quite good for
the density \# 1. For the density \# 2 it is rather unsatisfactory
for $x=0,$ which is explainable by the fact that in general there
are very few observations originating from the neighbourhood of
this point. Finally, we notice that the match is reasonably good
for the density \# 3, given the fact that it is difficult to
estimate, at the point $x=2.04,$ but is still unsatisfactory at
the point $x=0.$ The latter is explainable by the fact that there
are less observations originating from the neighbourhood of this
point. An increase in the sample size ($n=100$ and $n=200$) leads
to an improvement of the match between the theoretical and the
sample mean and standard deviation at the point $x=0$ for the
density \# 2, see Figure \ref{chifan} and Table \ref{chifantable},
however the results are still largely inaccurate for this point.
In essence similar conclusions were obtained for the density \# 3. These are not reported here.

\begin{table}[htb]
\begin{center}
\begin{tabular}{|c|c|c|c|c|c|}
\hline
$f$ & $h$ & $\hat{\mu}_1$ & $\hat{\mu}_2$ & $\hat{\sigma}_1$ & $\hat{\sigma}_2$\\
\hline
\# 1 & 0.24 & -0.046 & -0.093 & 0.953 & 1.127\\
\hline
\# 2 & 0.18 & -3.984 & -0.084 & 17.2 & 1.28\\
\hline
\# 3 & 0.25 & -0.768 & -0.141 & 4.03 & 1.63\\
\hline
\end{tabular}
\caption{\label{fantable} Sample means $\hat{\mu}_1$ and $\hat{\mu}_2$ and sample standard deviations $\hat{\sigma}_1$ and $\hat{\sigma}_2$ evaluated at $x=0$ and $x=0.92$ (densities \# 1 and \# 2) and $x=0$ and $x=2.04$ (the density \# 3).}
\end{center}
\end{table}

\begin{table}[htb]
\begin{center}
\begin{tabular}{|c|c|c|c|c|c|}
\hline
$n$ & $h$ & $\hat{\mu}_1$ & $\hat{\mu}_2$ & $\hat{\sigma}_1$ & $\hat{\sigma}_2$\\
\hline
100 & 0.17 & -1.33 & -0.015 & 9.89 & 1.31\\
\hline
200 & 0.15 & -1.02 & -0.015 & 6.36 & 1.58\\
\hline
\end{tabular}
\caption{\label{chifantable} Sample means $\hat{\mu}_1$ and $\hat{\mu}_2$ and sample standard deviations $\hat{\sigma}_1$ and $\hat{\sigma}_2$ evaluated at $x=0$ and $x=0.92$ for the density \# 2 for two sample sizes: $n=100$ and $n=200.$}
\end{center}
\end{table}

\begin{figure}[htb]
\setlength{\unitlength}{1cm}
\begin{minipage}{5.5cm}
\begin{picture}(5.5,4.0)
\epsfxsize=5.5cm\epsfysize=4cm\epsfbox{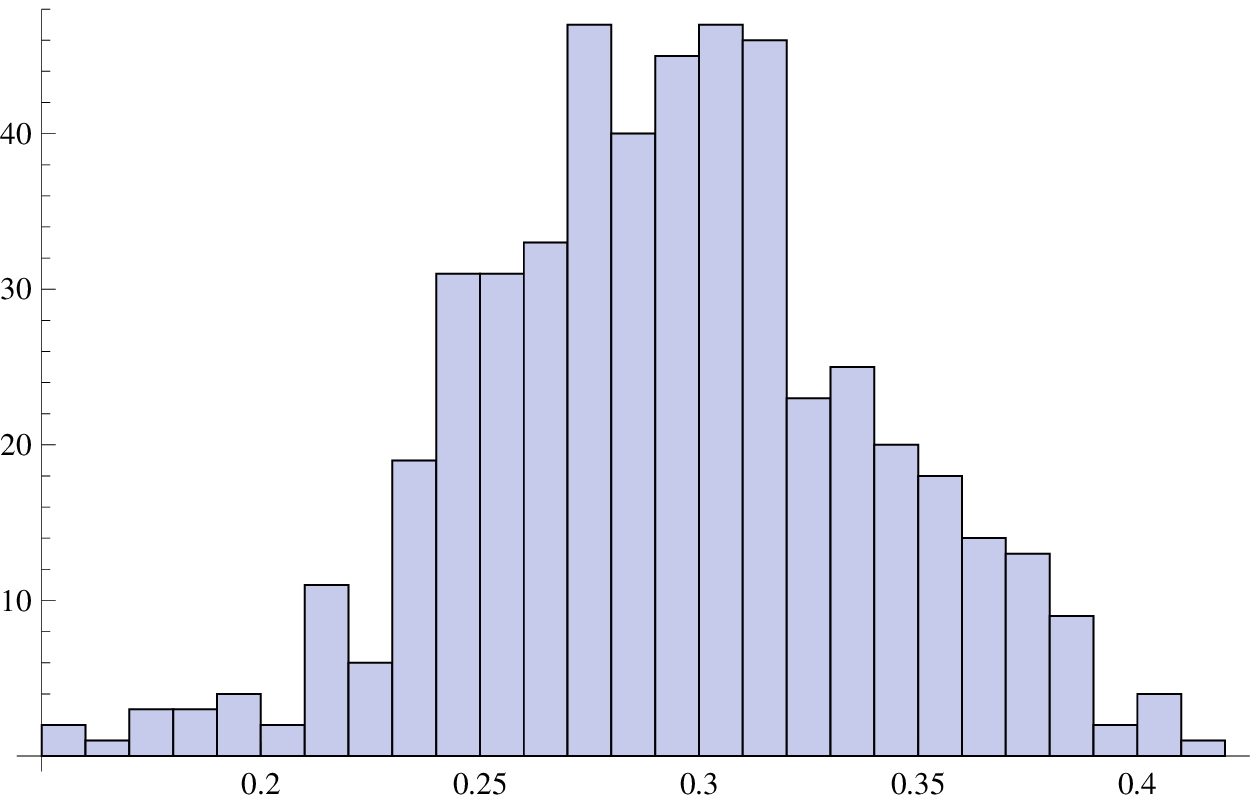}
\end{picture}
\end{minipage}
\hfill
\begin{minipage}{5.5cm}
\begin{picture}(5.5,4.0)
\epsfxsize=5.5cm\epsfysize=4cm\epsfbox{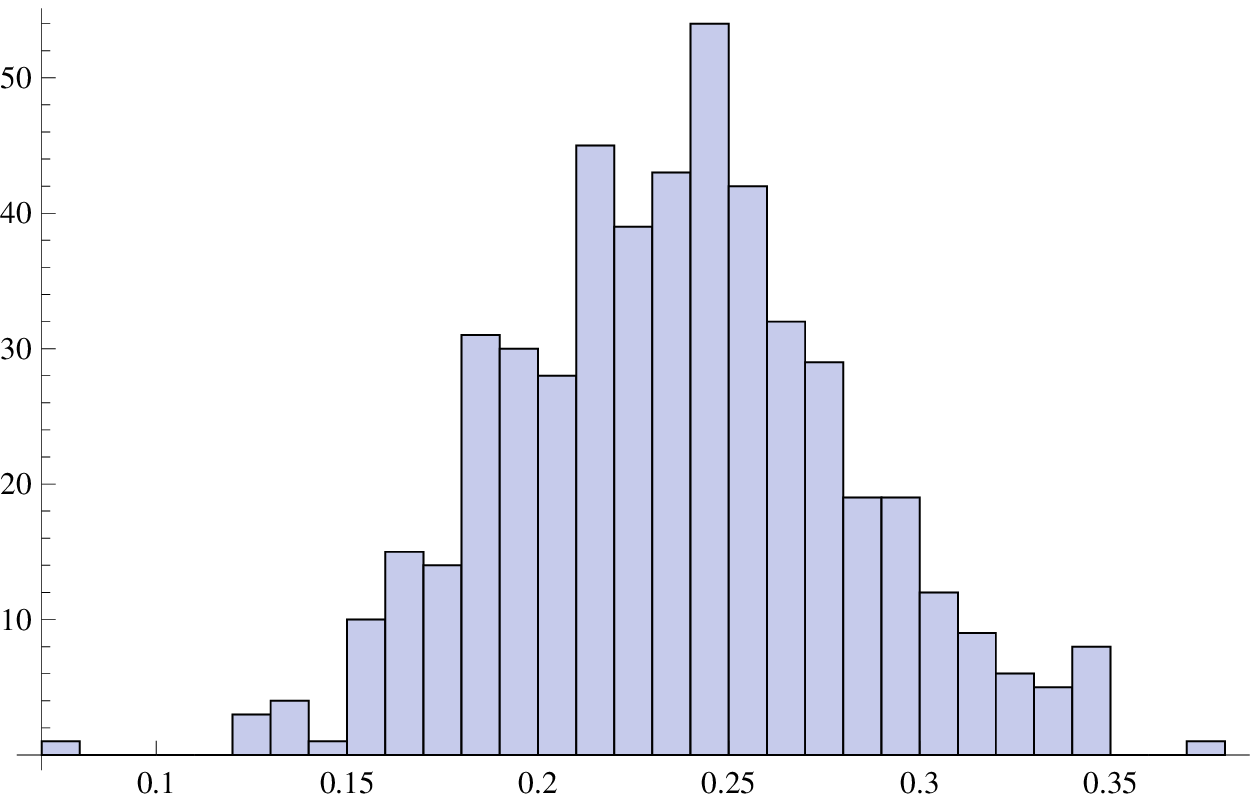}
\end{picture}
\end{minipage}
\setlength{\unitlength}{1cm}
\begin{minipage}{5.5cm}
\begin{picture}(5.5,4.0)
\epsfxsize=5.5cm\epsfysize=4cm\epsfbox{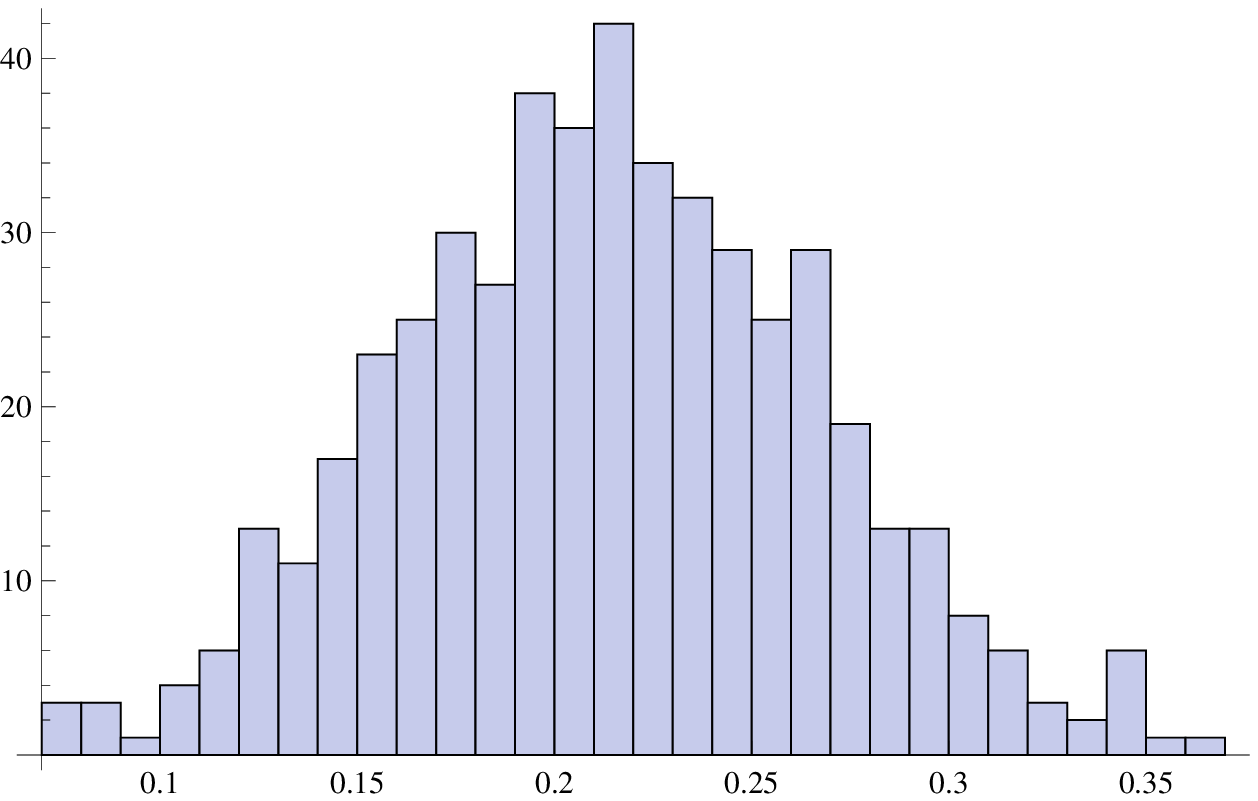}
\end{picture}
\end{minipage}
\hfill
\begin{minipage}{5.5cm}
\begin{picture}(5.5,4.0)
\epsfxsize=5.5cm\epsfysize=4cm\epsfbox{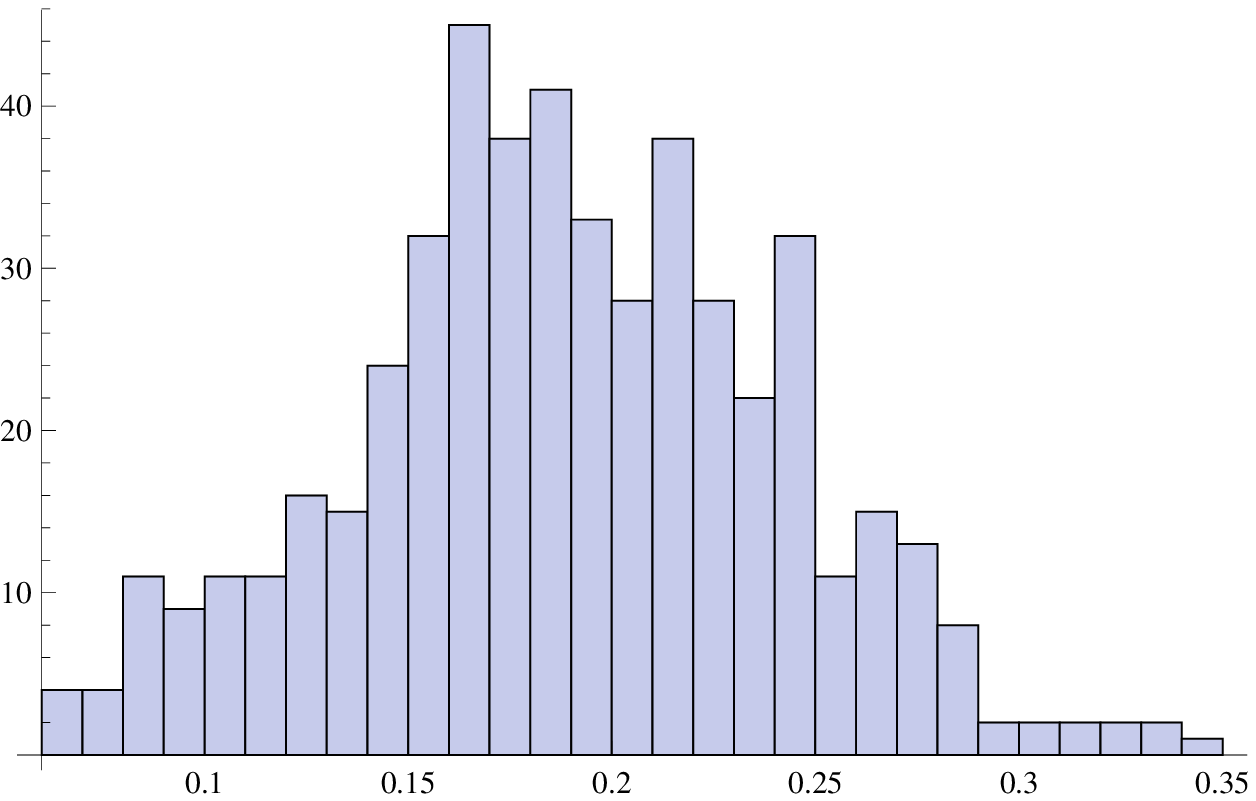}
\end{picture}
\end{minipage}
\begin{minipage}{5.5cm}
\begin{picture}(5.5,4.0)
\epsfxsize=5.5cm\epsfysize=4cm\epsfbox{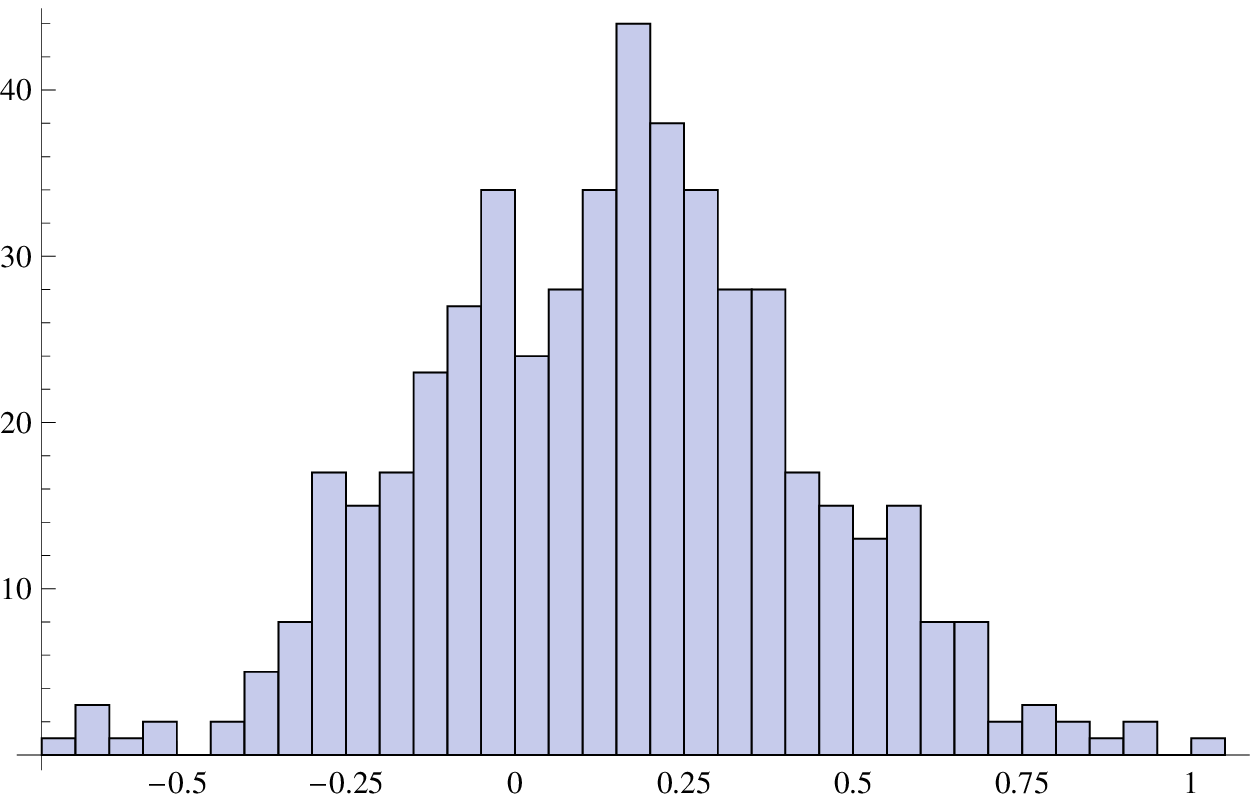}
\end{picture}
\end{minipage}
\hfill
\begin{minipage}{5.5cm}
\begin{picture}(5.5,4.0)
\epsfxsize=5.5cm\epsfysize=4cm\epsfbox{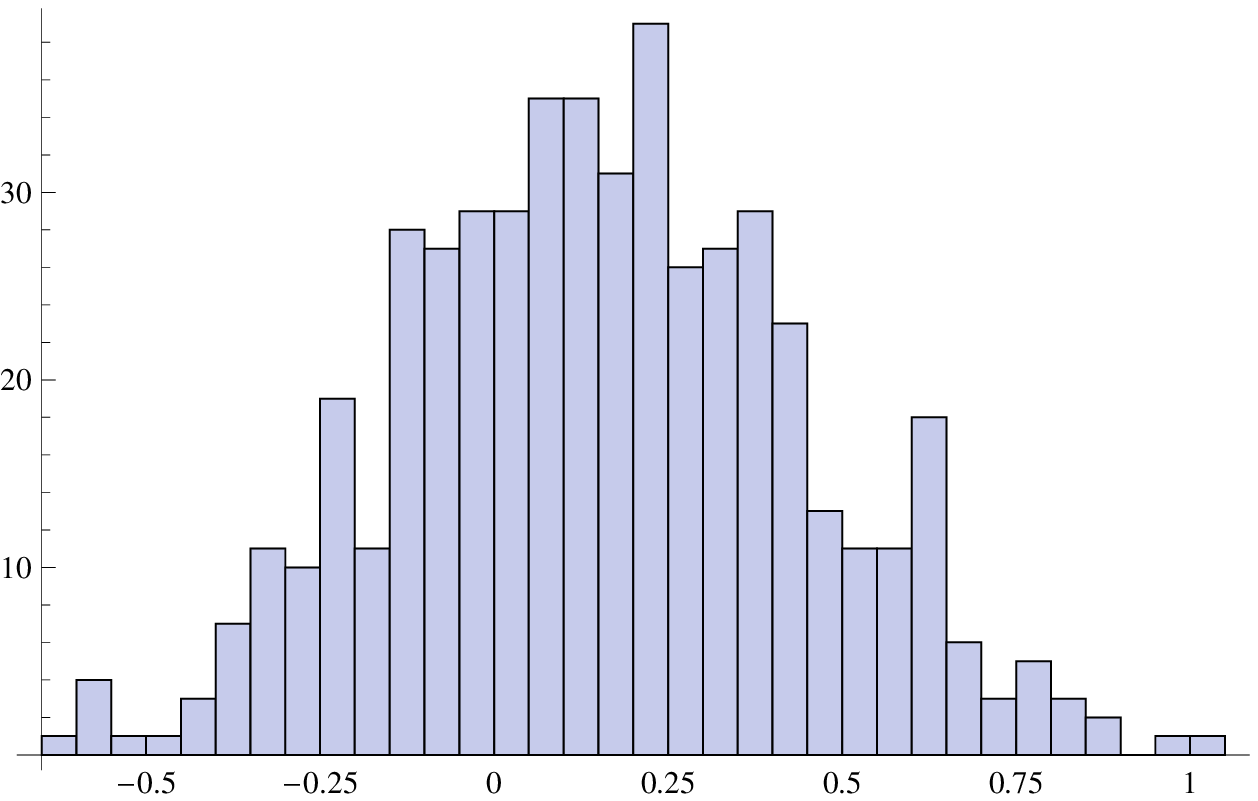}
\end{picture}
\end{minipage}
\caption{\label{highnoise} The histograms of $f_{nh}(x)$ for $x=0$ and $x=0.92$ for the density \# 1 for $n=50$ and three noise levels: $\operatorname{NSR}=100\%$ (top two graphs), $\operatorname{NSR}=400\%$ (middle two graphs) and $\operatorname{NSR}=1600\%$ (bottom two graphs).}
\end{figure}

\begin{figure}[htb]
\setlength{\unitlength}{1cm}
\begin{minipage}{5.5cm}
\begin{picture}(5.5,4.0)
\epsfxsize=5.5cm\epsfysize=4cm\epsfbox{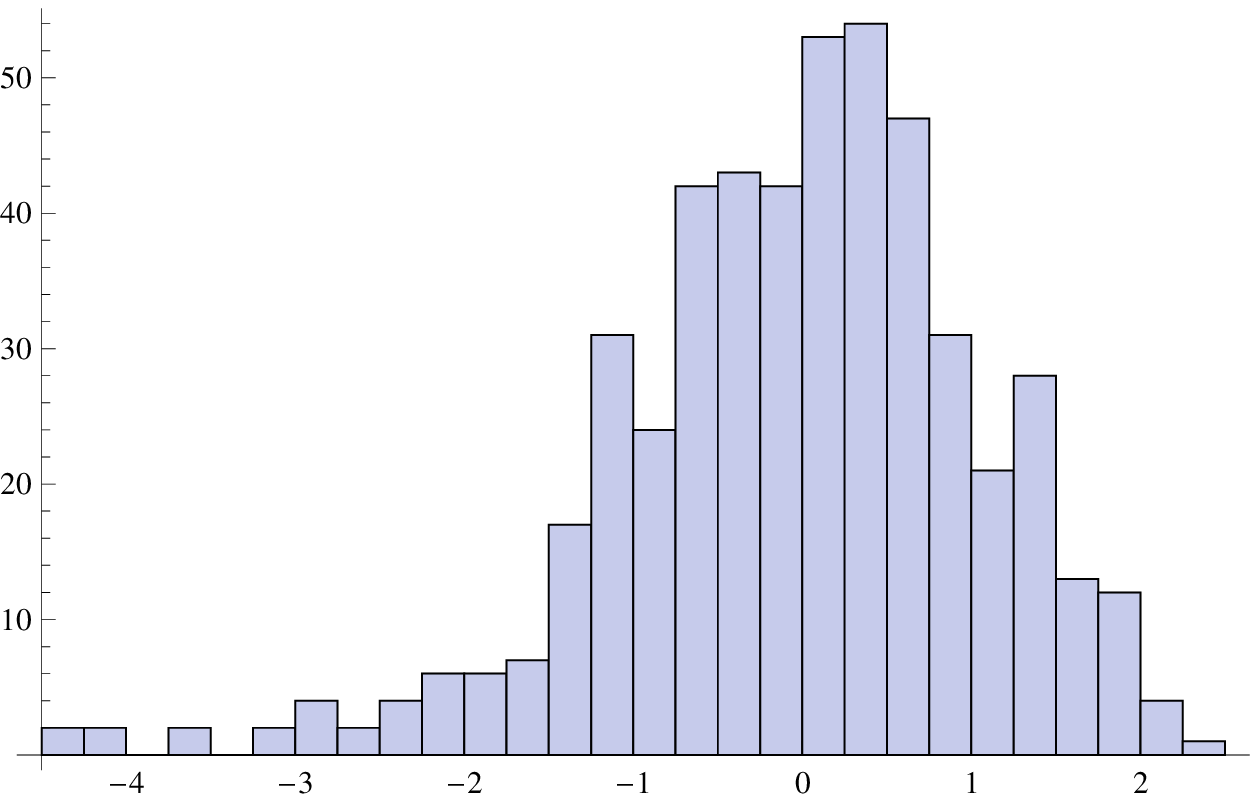}
\end{picture}
\end{minipage}
\hfill
\begin{minipage}{5.5cm}
\begin{picture}(5.5,4.0)
\epsfxsize=5.5cm\epsfysize=4cm\epsfbox{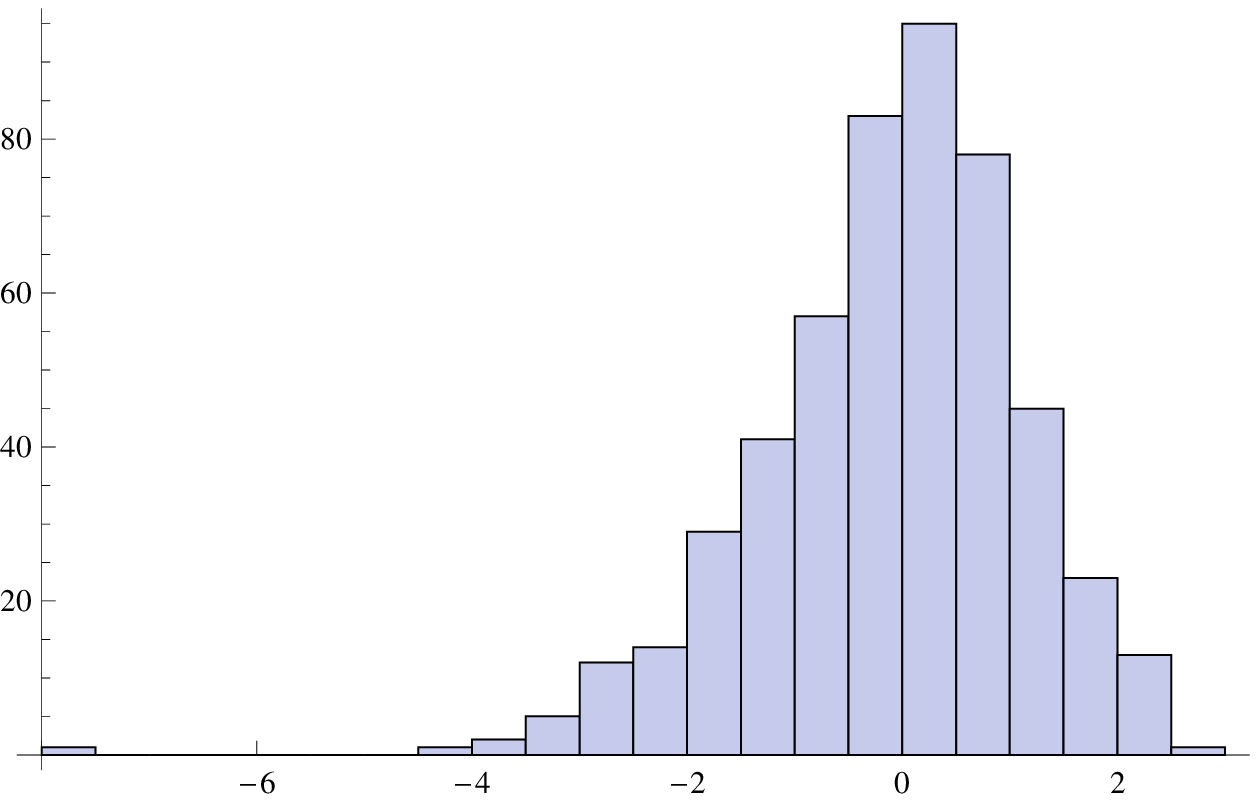}
\end{picture}
\end{minipage}
\setlength{\unitlength}{1cm}
\begin{minipage}{5.5cm}
\begin{picture}(5.5,4.0)
\epsfxsize=5.5cm\epsfysize=4cm\epsfbox{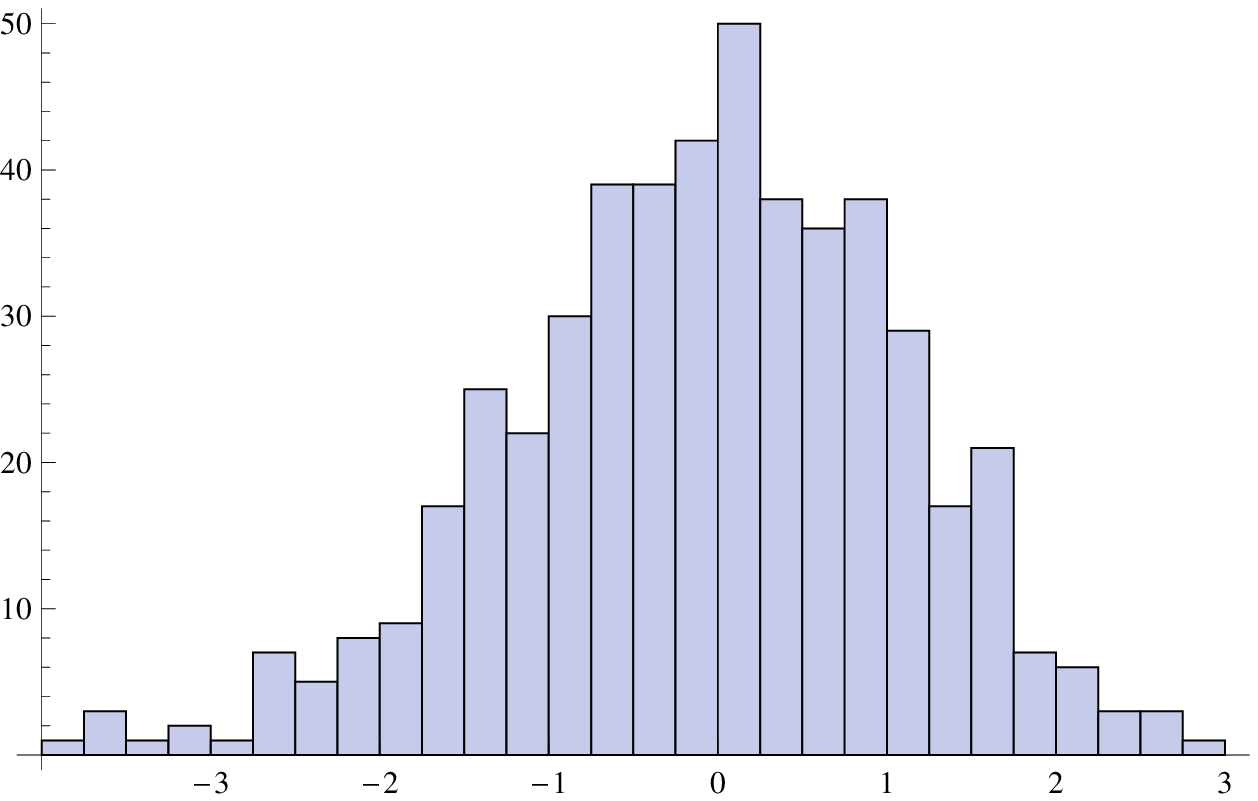}
\end{picture}
\end{minipage}
\hfill
\begin{minipage}{5.5cm}
\begin{picture}(5.5,4.0)
\epsfxsize=5.5cm\epsfysize=4cm\epsfbox{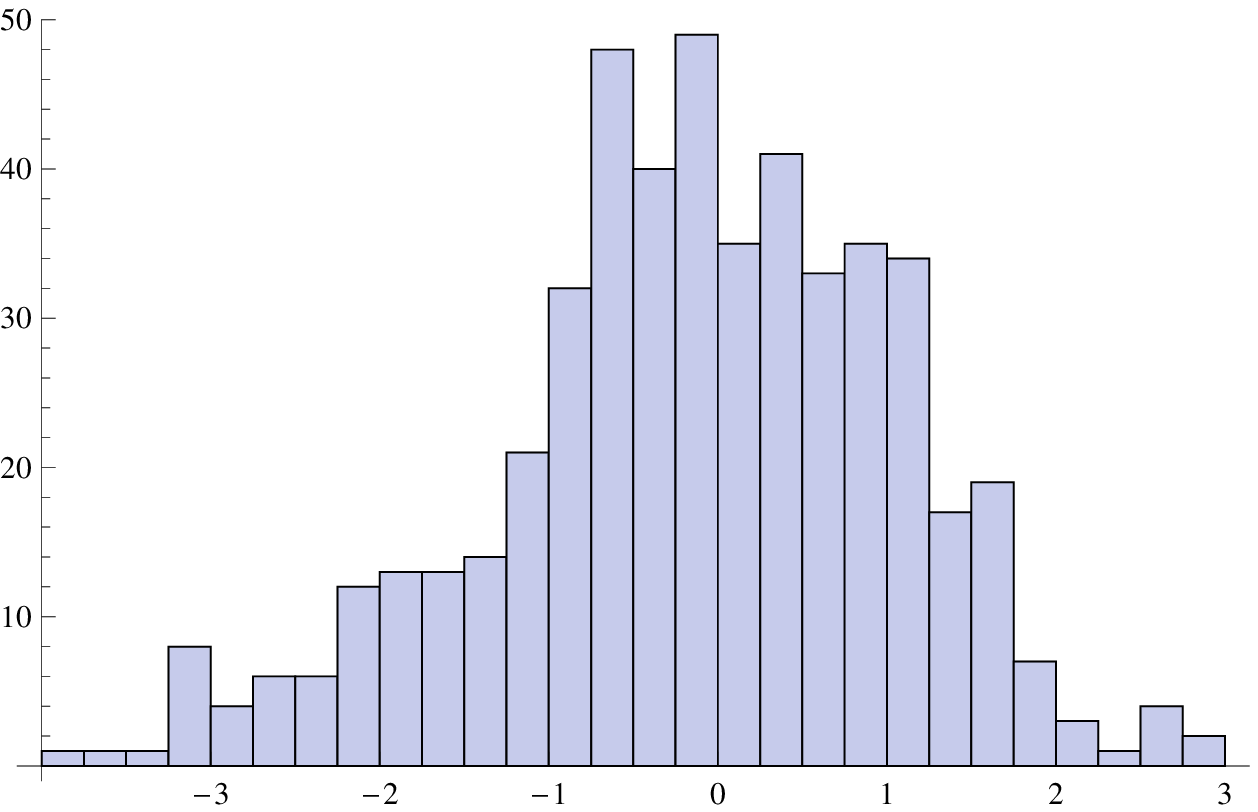}
\end{picture}
\end{minipage}
\begin{minipage}{5.5cm}
\begin{picture}(5.5,4.0)
\epsfxsize=5.5cm\epsfysize=4cm\epsfbox{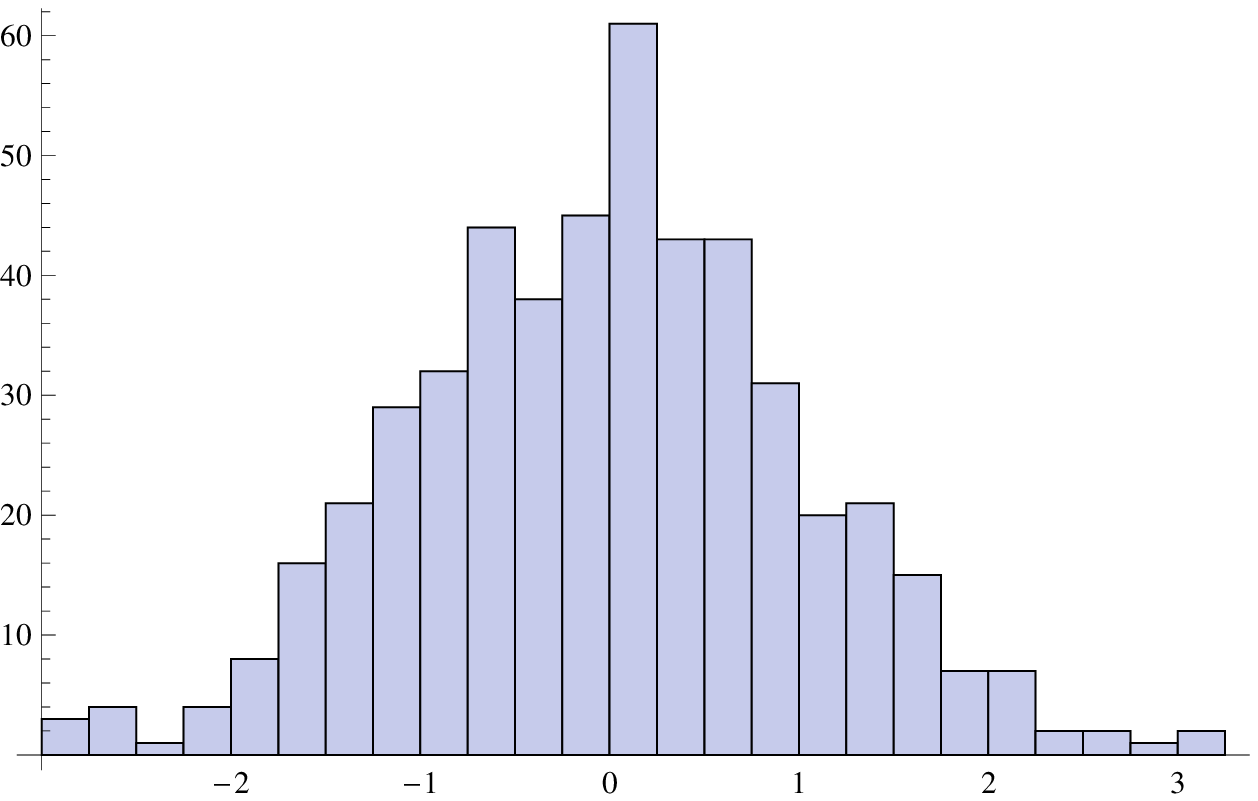}
\end{picture}
\end{minipage}
\hfill
\begin{minipage}{5.5cm}
\begin{picture}(5.5,4.0)
\epsfxsize=5.5cm\epsfysize=4cm\epsfbox{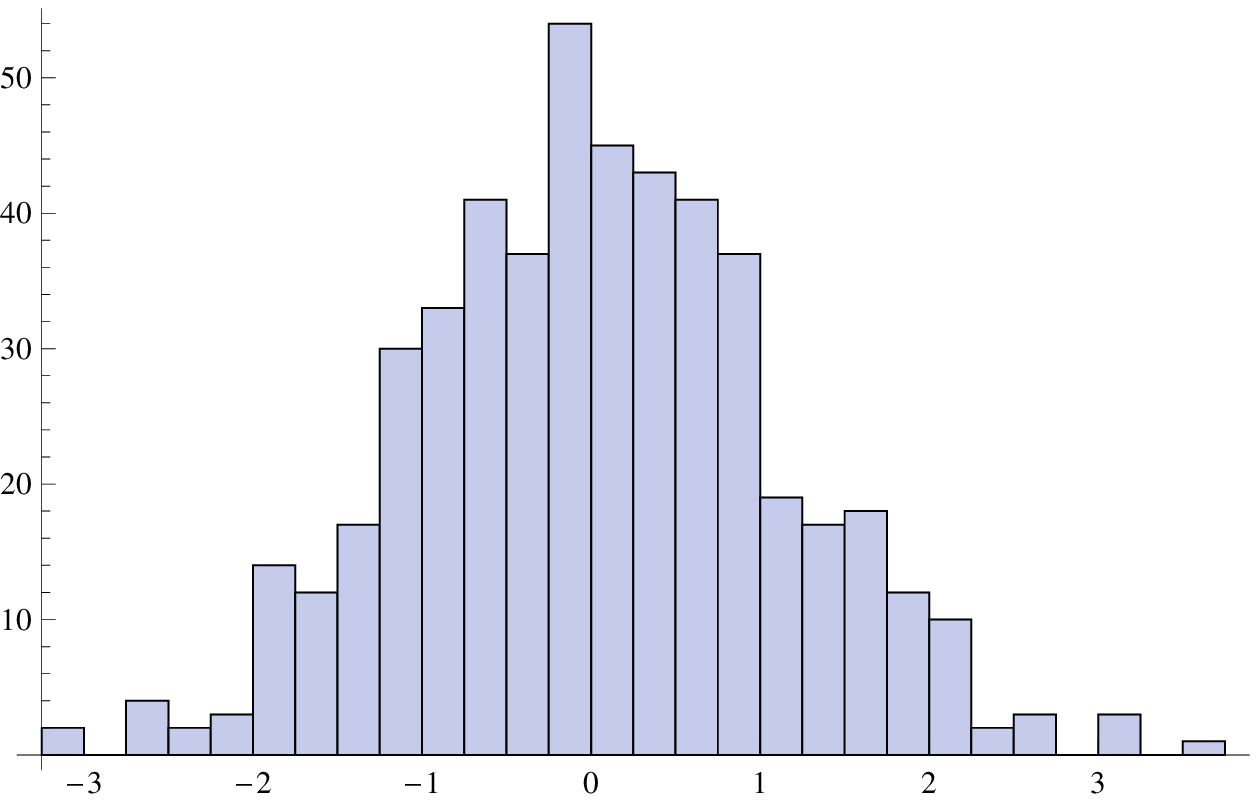}
\end{picture}
\end{minipage}
\caption{\label{fanhighnoise} The histograms of \eqref{fanexpr} for $x=0$ and $x=0.92$ for the density \# 1 for $n=50$ and three noise levels: $\operatorname{NSR}=400\%$ (top two graphs), $\operatorname{NSR}=400\%$ (middle two graphs) and $\operatorname{NSR}=1600\%$ (bottom two graphs).}
\end{figure}

Note that in all three models that we studied the noise level is
not high. We also studied the case when the noise level is very
high. For brevity we present the results only for the density \# 1
and for sample size $n=50.$ We considered three cases of the error
distribution: in the first case $Z\sim {\mathcal N}(0,1),$ in the
second case $Z\sim {\mathcal N}(0,2^2)$ and in the third case
$Z\sim {\mathcal N}(0,4^2).$ Notice that the $\operatorname{NSR}$
is equal to $100\%,$ $400\%$ and $1600\%,$ respectively. The
simulation results are summarised in Figures \ref{highnoise} and
\ref{fanhighnoise} and Tables \ref{tablehighnoise} and
\ref{fantablehighnoise}. We see that the sample standard deviation
and the corrected theoretical standard deviation are in better
agreement among each other compared to the low noise level case.
Also the histograms of the values of \eqref{fanexpr} look better.
On the other hand the resulting curves  $f_{nh}$ were not too
satisfactory when compared to the true density $f$ in the two
cases $Z\sim {\mathcal N}(0,1),$ and $Z\sim {\mathcal N}(0,2^2)$
(especially in the second case) and were totally unacceptable in
the case $Z\sim {\mathcal N}(0,4^2).$ This of course does not
imply that the estimator \eqref{fnh} is bad, rather the
deconvolution problem is very difficult in these cases.

\begin{table}[htb]
\begin{center}
\begin{tabular}{|c|c|c|c|c|c|c|c|}
\hline
$\operatorname{NSR}$ & $h$ & $\hat{\mu}_1$ & $\hat{\mu}_2$ & $\hat{\sigma}_1$ & $\hat{\sigma}_2$ & $\sigma$ & $\tilde{\sigma}$\\
\hline
100\% & 0.36 & 0.294 & 0.236 & 0.046 & 0.045 & 0.057 & 0.075\\
\hline
400\% & 0.59 & 0.214 & 0.189 & 0.053 & 0.053 & 0.046 & 0.076\\
\hline
1600\% & 0.89 & 0.150 & 0.156 & 0.279 & 0.289 & 0.251 & 0.342\\
\hline
\end{tabular}
\caption{\label{tablehighnoise} Sample means $\hat{\mu}_1$ and $\hat{\mu}_2$ and sample standard deviations $\hat{\sigma}_1$ and $\hat{\sigma}_2$ together with theoretical standard deviation $\sigma$ and corrected theoretical standard deviation $\tilde{\sigma}$ evaluated at $x=0$ and $x=0.92$ for the density \# 1 for three noise levels: $\operatorname{NSR}=100\%,$ $\operatorname{NSR}=400\%$ and $\operatorname{NSR}=1600\%$.}
\end{center}
\end{table}

\begin{table}[htb]
\begin{center}
\begin{tabular}{|c|c|c|c|c|c|}
\hline
$\operatorname{NSR}$ & $h$ & $\hat{\mu}_1$ & $\hat{\mu}_2$ & $\hat{\sigma}_1$ & $\hat{\sigma}_2$\\
\hline
100\% & 0.36 & -0.038 & -0.098 & 1.091 & 1.228\\
\hline
400\% & 0.59 & -0.079 & -0.134 & 1.155 & 1.193\\
\hline
1600\% & 0.89 & -0.015 & 0.035 & 1.027 & 1.086\\
\hline
\end{tabular}
\caption{\label{fantablehighnoise} Sample means $\hat{\mu}_1$ and $\hat{\mu}_2$ and sample standard deviations $\hat{\sigma}_1$ and $\hat{\sigma}_2$ of \eqref{fanexpr} evaluated at $x=0$ and $x=0.92$ for the density \# 1 for two noise levels: $\operatorname{NSR}=400\%$ and $\operatorname{NSR}=1600\%$.}
\end{center}
\end{table}

Finally, we mention that results qualitatively similar to the ones presented in this section were obtained for the kernel \eqref{sinckernel} as well. These are not reported here because of space restrictions.

\section{Discussion}
\label{conclusions}

In the simulation examples considered in Section \ref{simulations} for Theorem
\ref{thman}, we notice that the corrected theoretical asymptotic
standard deviation is always considerably larger than the sample
standard deviation given the fact that the noise level is not
high. We conjecture, that this might be true for the densities
other than \# 1, \# 2 and \# 3 as well in case when the noise
level is low. This possibly is one more explanation of the fact of
a reasonably good performance of deconvolution kernel density
estimators in the supersmooth error case for relatively small
sample sizes which was noted in \citet{wand}. On the other hand
the match between the sample standard deviation and the corrected
theoretical standard deviation is much better for higher levels of
noise. These observations suggest studying the asymptotic
distribution of the deconvolution kernel density estimator under
the assumption $\sigma\rightarrow 0$ as $n\rightarrow\infty,$ cf.\
\citet{delaigle0}, where $\sigma$ denotes the standard deviation
of the noise term.

Our simulation examples suggest that the asymptotic standard deviation evaluated via Theorem
\ref{thman} in general will not lead to an accurate approximation of the sample standard deviation, unless the
bandwidth is small enough, which implies that the corresponding sample size must be rather large. The latter is hardly
ever the case in practice. On the other hand, we have seen that in certain cases this poor approximation can be improved by using the left-hand side of \eqref{asnrm2} instead of the right-hand side. A perfect match is impossible to obtain given that we still neglect the remainder term in
\eqref{asnrm1}. However, even after the correction step, the corrected theoretical standard deviation still differs from the sample standard deviation considerably for small sample sizes and lower levels of noise. Moreover, in some cases the corrected theoretical standard deviation is even farther from the sample standard deviation than the original uncorrected version. The latter fact can be explained as follows:
\begin{enumerate}
\item It seems that both the theoretical and corrected theoretical standard deviation overestimate the sample standard deviation.
\item The value of the bandwidth $h,$ for which the match between the corrected theoretical standard deviation and the sample standard deviation become worse, belongs to the range where the corrected theoretical standard deviation is larger than the theoretical standard deviation. In view of item 1 above, it is not surprising that in this case the theoretical value turns out to be closer to the sample standard deviation than the corrected theoretical value.
\end{enumerate}

The consequence of the above observations is that a naive attempt
to directly use Theorem \ref{thman}, e.g.\ in the construction of
pointwise confidence intervals, will lead to largely inaccurate
results. An indication of how large the contribution of the
remainder term in \eqref{asnrm1} can be can be obtained only after a thorough
simulation study for various distributions and sample sizes, a
goal which is not pursued in the present note. From the three
simulation examples that we considered, it appears that the
contribution of the remainder term in \eqref{asnrm1} is quite
noticeable for small sample sizes. For now we would advise to use
Theorem \ref{thman} for small sample sizes and lower noise levels
with caution. It seems that the similar cautious approach is
needed in case of Theorem \ref{thmanfan} as well, at least for
some values of $x.$

Unlike for the ordinary smooth case, see \citet{bissantz}, there
is no study dealing with the construction of uniform confidence
intervals in the supersmooth case. In the latter paper a better
performance of the bootstrap confidence intervals was demonstrated
in the ordinary smooth case compared to the asymptotic confidence
bands obtained from the expression for the asymptotic variance in
the central limit theorem. The main difficulty in the supersmooth
case is that the asymptotic distribution of the supremum distance
between the estimator $f_{nh}$ and the true density $f$ is
unknown. Our simulation results seem to indicate that the
bootstrap approach is more promising for the construction of
pointwise confidence intervals than e.g.\ the direct use of
Theorems \ref{thmanfan} or \ref{thman}. Moreover, the simulations
suggest that at least Theorem \ref{thman} is not appropriate when
the noise level is low.

\end{document}